\documentclass[authoryear,preprint,3p,review,12pt]{elsarticle}
 \usepackage{graphics}
 \usepackage{rotating}
 \usepackage{color}
\usepackage{amssymb}
 \usepackage{amsthm}

\usepackage{setspace}

\journal{Icarus}

\begin{document}

\begin{frontmatter}

%% Title, authors and addresses

%% use the tnoteref command within \title for footnotes;
%% use the tnotetext command for the associated footnote;
%% use the fnref command within \author or \address for footnotes;
%% use the fntext command for the associated footnote;
%% use the corref command within \author for corresponding author footnotes;
%% use the cortext command for the associated footnote;
%% use the ead command for the email address,
%% and the form \ead[url] for the home page:
%%
%% \title{Title\tnoteref{label1}}
%% \tnotetext[label1]{}
%% \author{Name\corref{cor1}\fnref{label2}}
%% \ead{email address}
%% \ead[url]{home page}
%% \fntext[label2]{}
%% \cortext[cor1]{}
%% \address{Address\fnref{label3}}
%% \fntext[label3]{}

%\title{Detection of benzene ice in Titan's fall southern stratospheric polar cloud with Cassini/CIRS}
\title{Study of Titan's fall southern stratospheric polar cloud composition with Cassini/CIRS: detection of 
benzene ice}

%% use optional labels to link authors explicitly to addresses:
%% \author[label1,label2]{<author name>}
%% \address[label1]{<address>}
%% \address[label2]{<address>}

\author[LESIA]{S. Vinatier\corref{cor1}}
\ead{sandrine.vinatier@obspm.fr}
\cortext[cor1]{Corresponding author}

\author[IPAG]{B. Schmitt}
\author[LESIA]{B. B\'ezard}
\author[GSMA]{P. Rannou}
\author[Villebon]{C. Dauphin}
\author[UTRECHT]{R. de Kok}
\author[GSFC]{D. E. Jennings}
\author[GSFC]{F. M. Flasar}

%\author[LMD]{S\'ebastien Lebonnois}
%\author[Bristol]{Nick Teanby}
%\author[Maryland,GSFC]{Richard Achterberg}
%\author[GSFC]{Nicolas Gorius}
%\author[GSFC]{Andrei Mamoutkine}
%\author[GSFC]{Ever Guandique}
%\author[LISA]{Antoine Jolly}
%\author[GSFC]{Don Jennings}
%\author[GSFC]{Mike Flasar}

%\author[NASA]{Carrie M. Anderson} 
%\author[SRON]{Remco de Kok}
%\author[MARYLAND,NASA]{Robert E. Samuelson}

\address[LESIA]{LESIA, Observatoire de Paris, PSL Research University, CNRS, Sorbonne Universit\'es, UPMC Univ. Paris 06, Univ. Paris Diderot, Sorbonne Paris Cit\'e, 5 place Jules Janssen, 92195 Meudon, France}
\address[IPAG]{Universit\'e Grenoble Alpes, CNRS, Institut de Plan\'etologie et d'Astrophysique de Grenoble (IPAG), France}
\address[GSMA]{GSMA, UMR CNRS 6089, Univ. de Reims Champagne-Ardenne, France}
\address[Villebon]{Institut Villebon - Georges Charpak, D\'epartement de Physique - UFR Sciences, Universit\'e Paris Sud, Bat. 490, rue Hector Berlioz, 91400 Orsay, France}
\address[UTRECHT]{Department of Physdical Geography, Utrecht University, P.O. Box 80115, 3508 TC Utrecht, The Netherlands}
\address[GSFC]{NASA/Goddard Space Flight Center, Code 693, Greenbelt, MD 20771, USA.}

\address{}

\begin{abstract}
%% Text of abstract
\textcolor{black}{We report the detection of a spectral signature observed at 682 cm$^{-1}$ by the Cassini Composite Infrared Spectrometer (CIRS) 
 in nadir and limb geometry observations of Titan's southern stratospheric polar region in the middle of southern fall,
 while stratospheric temperatures are the coldest since the beginning of the Cassini mission. 
In the same period, many gases observed in CIRS spectra (C$_2$H$_2$, HCN, C$_4$H$_2$, C$_3$H$_4$, HC$_3$N and C$_6$H$_6$) are highly enriched in the stratosphere at high southern latitude due to the air subsidence of the global atmospheric circulation and 
some of these molecules condense at much higher altitude than usually 
observed for other latitudes.
The 682 cm$^{-1}$ signature, which is only observed below an altitude of 300-km, is at least partly attributed to the benzene (C$_6$H$_6$) 
ice $\nu_{4}$ C-H bending mode. 
While we first observed it} in CIRS nadir spectra of the southern polar region in early 2013, we focus here on the study of nadir data acquired in May 2013, which have a more favorable observation geometry. We derived the C$_6$H$_6$ ice mass mixing ratio in 5$^{\circ}$ latitude bins from the south pole to 65$^{\circ}$S
and infer the C$_6$H$_6$ cloud top altitude to be located deeper with increasing distance from the pole. 
We additionally analyzed limb data acquired in March 2015, which were the first limb dataset available after the May 2013 nadir observation, 
in order to infer a vertical profile of its mass mixing ratio in the 0.1 - 1 mbar region (250 - 170 km). 
We derive an upper limit of $\sim$1.5 $\mu$m for the equivalent radius of pure C$_6$H$_6$ 
ice particles from the shape of the observed emission band, \textcolor{black}{which is consistent with our estimation of the 
ice particle size from condensation growth and sedimentation timescales.}
We compared the ice mass mixing ratio with the haze mass mixing ratio inferred in the same region from the continuum emission of CIRS spectra, and derived that the 
haze mass mixing ratios are $\sim$30 times larger than the C$_6$H$_6$ ice mass mixing ratios for all observations.
Several other unidentified signatures are observed near 687 and 702 cm$^{-1}$ and possibly 695 cm$^{-1}$, which could also be due to ice spectral signatures 
as they are observed in the deep stratosphere at pressure levels similar to the C$_6$H$_6$ ice ones. We could not reproduce these signatures
with pure nitrile ices \textcolor{black}{(HCN, HC$_3$N,CH$_3$CN, C$_2$H$_5$CN and C$_2$N$_2$)} spectra available in the literature except the 695 cm$^{-1}$ feature that could possibly be due to 
C$_2$H$_3$CN ice. \textcolor{black}{From this tentative detection, we derive the corresponding C$_2$H$_3$CN ice} mass mixing ratio profile and also inferred an upper limit of its gas volume mixing 
ratio of 2$\times$10$^{-7}$ at 0.01 mbar at 79$^{\circ}$S in March 2015.

\end{abstract}

\begin{keyword}
Titan, atmosphere \sep Infrared observations \sep Atmospheres, structure \sep Atmospheres, composition

%% keywords here, in the form: keyword \sep keyword

%% MSC codes here, in the form: \MSC code \sep code
%% or \MSC[2008] code \sep code (2000 is the default)
\end{keyword}

\end{frontmatter}

%\linenumbers

{\bf Highlights:}

\begin{itemize}

\item \textcolor{black}{We studied Titan's fall southern stratospheric polar cloud composition with Cassini/CIRS.}

\item \textcolor{black}{We have detected C$_6$H$_6$ ice through its 682 cm$^{-1}$ $\nu_4$ C-H bending mode.}

\item \textcolor{black}{We derive an upper limit of 1.5 $\mu$m for the equivalent radius of C$_6$H$_6$ ice particles.}

\item \textcolor{black}{Vertical and spatial distribution of C$_6$H$_6$ ice mass mixing ratio is derived in the South polar cloud.}

\item \textcolor{black}{We investigate potential ice candidates for other detected spectral features.}

\end{itemize}

%% main text
\section{Introduction}\label{intro}

Since the northern spring equinox in August 2009, Titan's stratospheric and mesospheric thermal field 
and minor species mixing ratio distributions have experienced very strong seasonal changes.
Analysis of the Cassini Composite Infrared Spectrometer (CIRS) 
showed that the dynamical descending branch predicted by General Circulation Models \citep{Lebonnois_2012,Newman_2011,Larson_2014}
was observed at the south pole for the first time in June 2010 through
its adiabatic heating that warmed up the mesosphere around 400 km and through enhancement of haze confined at latitudes 
higher than 80$^{\circ}$S \citep{Teanby_2012,Vinatier_2015}. 
This descending branch also brought molecular enriched air from the upper atmosphere, where molecules are formed, 
towards deeper levels, and the first molecular enhancements were observed above the South pole in June 2011 above 400 km \citep{Teanby_2012,Vinatier_2015,Coustenis_2016}.
As southern autumn progressed these enhancements were observed at lower altitude due to the 
downward transport of air by the descending branch. However, while temperature in the upper atmosphere 
was expected to increase by adiabatic heating due to the predicted reinforcement of the descending branch vertical velocity, 
an unexpected thermal cooling was observed in January 2012 in the 350-500 km range \citep{de_Kok_2014,Vinatier_2015,Teanby_2017,Vinatier_2016}. 
This cooling is partly due to the radiative cooling by the highly enriched molecules at high altitude, which exceeds 
the adiabatic heating due to the descending branch \cite{Teanby_2017}. Additionally, another factor explaining the temperature decrease is
the decrease of solar flux during southern autumn. This results in a net cooling of the high southern 
latitudes, with observed temperatures as low as 115 K in the deep stratosphere \citep{Achterberg_2014}, which leads to condensation 
of gas at higher altitude than usually observed for other latitudes.
HCN ice was detected at 300 km in June 2012 from the Visual and Infrared 
Mapping Spectrometer (VIMS) limb observations \citep{de_Kok_2014}, coincident with the stratospheric polar cloud observed since May 2012 by the Cassini Imaging Science Subsystem \citep{West_2016} located at the same altitude and with a similar  horizontal extent ($\sim$600-900 km). This suggests that HCN ice could be an important component of the autumn south polar cloud. Additionally, in July 2012, CIRS first observed the 220 cm$^{-1}$ emission feature, attributed to condensates \citep{Coustenis_1999,de_Kok_2007b,de_Kok_2008}, at the south pole (while it was not observed in February 2012, \cite{Jennings_2012}). All these independent observations suggest a rapid formation of the stratospheric polar cloud between February and May 2012.
In the present study, we investigate the composition of the south polar cloud using CIRS nadir and 
limb observations in the 600-1400 cm$^{-1}$ wavenumber range in May 2013 (from nadir viewing) and 
in March 2015 (from limb viewing). We present here the detection 
of the $\nu_{4}$ benzene (C$_6$H$_6$) ice C-H bending vibration band at 682 cm$^{-1}$ \citep{Bertie_2004} and derive spatial constraints of its 
mass mixing ratio in May 2013 and its vertical extent in March 2015.
 
Section 2 describes observations used in this study. The retrieval method and the inferred temperature and C$_6$H$_6$ gas volume mixing ratio
profiles are presented in Section 3 and 4, respectively. Section 5 and 6 focus on the detection and retrieval of the C$_6$H$_6$
ice mass mixing ratio, respectively. Our results are discussed in Section 7.

\section{Observations}\label{section_obs}

CIRS observes the Titan thermal emission in the 
ranges 10 - 100 cm$^{-1}$ with focal plane 1 (FP1), 600 - 1100 cm$^{-1}$ with focal plane 3 (FP3)
and 1100 - 1500 cm$^{-1}$ with focal plane 4 (FP4). We focus here on spectra acquired by FP3 and FP4. Both focal planes are each composed 
of a linear 10 adjacent detectors array, each detector having a 0.275$\times$0.275 mrad field-of-view. 
During a limb observation, the projected FP3 and FP4 detector arrays are positioned perpendicular to the surface
so that each detector probes a different altitude range in the 10-0.001 mbar (100-500 km) region, above a given latitude with a typical 30 km vertical resolution, 
comparable to the pressure scale height ($\sim$ 40 km in the stratosphere). CIRS also acquires spectra with nadir geometry viewing, 
in sequences providing the global 
horizontal mapping during a given flyby, albeit probing the 10-1 mbar region ($\sim$ 100-150 km) with poor or no 
vertical resolution. 
Our study is based on analysis of CIRS nadir spectra acquired at 2.8 cm$^{-1}$ spectral resolution and limb spectra acquired at 
0.5 cm$^{-1}$ resolution.

	\subsection{Nadir observations in May 2013}

Between September 2012 (T86) and October 2014 (T106), Cassini orbits were highly inclined, with inclination higher than 
40$^{\circ}$ relative to the Saturn equatorial plane.
Such inclined orbits are the most suitable to map Titan's poles in nadir observing mode, while simultaneous CIRS limb observations of these
regions are impossible. In the present study, we chose to focus on data 
acquired in May 2013 (Titan's flyby T91) corresponding to the Cassini most inclined orbit in the 
period September 2012 - October 2014. Our selected May 2013 nadir observations were acquired during a flyby with an 
orbit inclination higher than 60$^{\circ}$. The only higher inclined orbit of the mission occurred in January 2017, a few months 
before the end of the mission. 

In order to derive information on the spatial distribution of the south polar cloud, we used nadir observations of the south polar region 
acquired with the smallest possible emission angle in order to reduce mixing of thermal emission from several latitudes along the line-of-sight, as the temperature latitudinal gradient near the South pole in autumn is strong. 

In order to increase signal-to-noise ratio, we averaged observed nadir spectra in latitudinal bins of 5$^{\circ}$ wide and in emission angle
bins of $\sim$10-20$^{\circ}$ wide with the smallest possible mean emission angle, assuming no longitudinal variations in each 5$^{\circ}$ bin.

Our spectra selections take into account the 4.1$^{\circ}$ offset of the stratospheric rotation axis relative to the solid body 
rotation axis \citep{Achterberg_2008} and its longitudinal drift of 9.15$^{\circ}$ per year 
in the sun-fixed frame derived by \cite{Achterberg_2011} from data acquired up to the northern spring equinox. In other words, 
the direction of the stratospheric rotation axis seems to be stationary in the stellar-fixed frame and we define our latitude from this axis.  
 At the equinox, in August 2009, \cite{Achterberg_2011} derived an offset azimuth $\sim$95$^{\circ}$ W from the subsolar longitude. 
From these results, we estimated that the north pole offset pointed towards a direction $\sim$130$^{\circ}$ W from 
the subsolar longitude for our observations in May 2013. 
    
The sub spacecraft latitude and longitude of observations used in this study were about 48$^{\circ}$S and
250$^{\circ}$W, respectively. 

%We assumed no longitudinal variations in each 5$^{\circ}$ bin. In order to increase the signal-to-noise ratio, 
%we averaged nadir spectra acquired at several longitudes in each latitudinal bin and with emission angles included in a range of $\sim$10-20$^{\circ}$ 
%wide with the smallest possible mean emission angle.

Table 1 gives characteristics of our selected spectra for each 5 degree latitudinal 
bin. Spectra were extracted from the 
v4.3.1 calibration version. Observed FP4 averaged spectra are displayed in Figure \ref{spe_FP4}, while observed FP3 averages 
are displayed in Figure \ref{spe_FP3}.

	\subsection{Limb observations in March 2015}
After February 2012, when the south polar stratospheric composition was probed with CIRS limb observations \citep{Teanby_2012, Vinatier_2015}, 
Cassini's high orbit inclinations prevented us to observe poles with limb viewing geometry till January 2015. 
We used here the first limb observation
of the south polar region acquired at 0.5 cm$^{-1}$ spectral resolution after our May 2013 nadir selection. 
These limb spectra were acquired in March 2015, during Titan flyby T110 (see their characteristics in Table 2).
In order to probe the deep stratospheric temperatures, we combined our limb spectra analysis with a nadir spectra average 
acquired at 3 cm$^{-1}$ spectral resolution in December 2014, during flyby T107, assuming that no temporal 
variations occurred in the deep stratosphere (around 10 mbar) between December 2014 and March 2015 
($\sim$ 6 Titan's days), which is justified regarding the radiative relaxation time of $\sim$10$^7$s ($\sim$ 36 Titan's days) 
at 10 mbar derived by \cite{Bezard_2017} for equatorial temperature (radiative relaxation time would be 
longer for the observed colder polar temperatures assuming comparable atmospheric composition at 10 mbar).

\section{Temperature profiles in the south polar region}   \label{section_profT}

	\subsection{Retrieval method} 

\begin{figure}[p] 
	\centering

	 \includegraphics[scale=0.7]{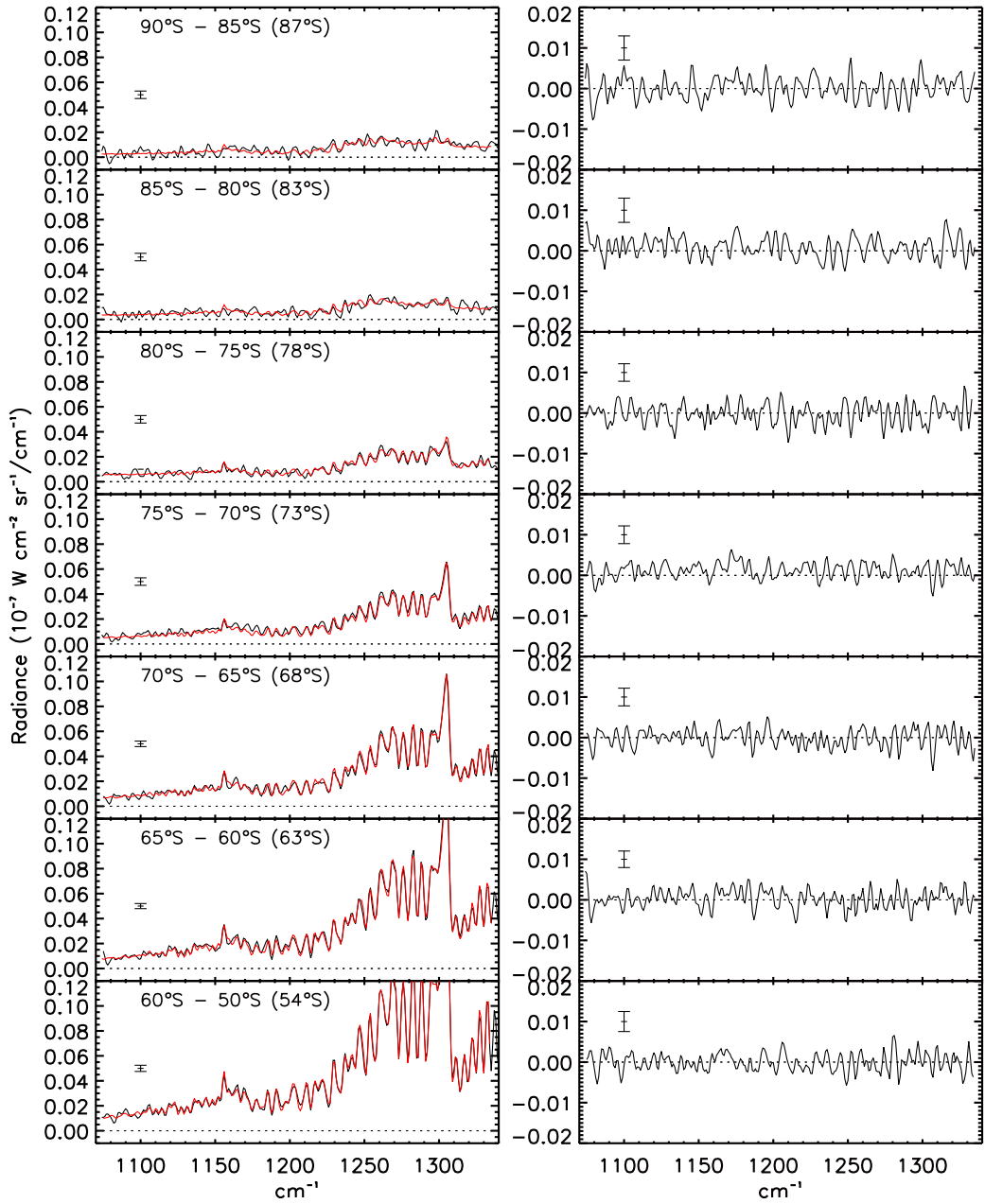}
         \caption{Left: CIRS spectra (black line) of the CH$_4$ $\nu_4$ band compared to the calculated spectra (red) 
	 from 87$^\circ$S to 54$^\circ$S (upper to lower panel) for the nadir observation of May 2017. Latitude ranges used for
	 the averages are given for each observed spectrum (with the mean latitude in brackets). Radiance 
	 scale is the same for all latitudes in order 
	 to emphasize the latitudinal variations of the CH$_4$ radiance, which monitors the meridonal thermal changes
	 as CH$_4$ mixing ratio is supposed to be constant. 
         Right: Corresponding residuals (observed spectrum minus calculated 
	 spectrum) with 1-$\sigma$
	 error bars. 
		 }	
	\label{spe_FP4}
	\end{figure}

Intensities of molecular thermal emission bands displayed in Figures \ref{spe_FP4} and \ref{spe_FP3} strongly depend
on temperature.	We therefore first derived the temperature vertical profile by
fitting the $\nu_4$ CH$_4$ band at 1305 cm$^{-1}$ (see Fig. \ref{spe_FP4} for May 2013 nadir observations), assuming a CH$_4$ volume mixing ratio 
constant with altitude and latitude and equal to 1.48\% \citep{Niemann_2010}. We used a linear constrained inversion algorithm described 
by \cite{Vinatier_2007a, Vinatier_2015} to retrieve simultaneously temperature and haze optical depth.
A first estimation of the haze optical depth was inferred from the 1070-1110 cm$^{-1}$ region, but as it was poorly constrained 
at very high latitude, because of the poor signal-to-noise ratio due to low stratospheric temperature, we 
corrected it in a second step by fitting the continuum of the 640-660 cm$^{-1}$ spectral region of the FP3 focal plane
(see Section \ref{section_gaz_retrieval}) in order to obtain the best fit of the FP3 continuum. We used the aerosol spectral dependence derived by \cite{Vinatier_2012}.
 
Our inversion method needs a priori profiles for both temperature and aerosol optical depth. For analysis of May 2013 nadir spectra, these a priori profiles were similar to those derived from limb observations of February 2012 \citep{Vinatier_2015}, which were constrained
in the  5-0.001 mbar range ($\sim$150-500 km) with a vertical resolution of $\sim$40 km. 
These vertical profiles of the south polar region are the closest in time to the May 2013 nadir observation 
as no limb data of the south pole were acquired between February 2012 and January 2015 because of the high inclined Cassini's orbits. 
For nadir geometry, thermal emission is integrated along the line-of-sight and,
for a given wavelength, the region of maximum emission is localized where opacity is close to 1. Nevertheless, 
some vertical information can be retrieved by simultaneously fitting the 
methane P-branch (wavenumbers lower than 1290 cm$^{-1}$), which probes the deep stratosphere in the 0.5-20 mbar, and the 
Q-branch (at 1305 cm$^{-1}$), which is more opaque and probes lower pressure levels, typically in the 0.05-1 mbar region.  
Then, by fitting the P and Q-branches of the CH$_4$ $\nu_4$ band, we were able to derive information \textcolor{black}{on the temperature profile} 
from 0.05 to 20 mbar.

The thermal profiles and haze optical depth derived from the May 2013 78$^{\circ}$S nadir observations were used as a priori of the retrieval 
of combined nadir and limb data acquired in December 2014 and March 2015, respectively. Usually, in order to obtain the best fit 
of the $\nu_4$ CH$_4$ band in limb spectra, we have to apply a shift on the altitude of line-of-sights extracted from the CIRS database. 
As explained in \cite{Vinatier_2007a}, this shift is due to the CIRS navigation pointing error and/or the calculated 
pressure/altitude grid from hydrostatic equilibrium based on our a priori thermal profile outside the
regions probed by CIRS. For our previous limb data analysis, we applied shifts usually varying from -30 km to +30 km
with error bars lower than 5 km \citep{Vinatier_2010a, Vinatier_2015}. For the limb data used in this study, because of smaller 
signal-to-noise due to lower temperatures than usually observed, the shift is poorly constrained and we derived best fits of the $\nu_4$ CH$_4$ band with a lower limit of +15 km (at 3-$\sigma$), a best fit value of +40 km, and no real constraint on its upper limit. We thus applied a
+40 km vertical shift on all line-of-sight altitudes of our selected limb spectra.

        \subsection{Results} 

Calculated spectra corresponding to the best fit of the observed nadir spectra acquired in May 2013  
are displayed in Figure \ref{spe_FP4} (left panel), with their corresponding residuals 
(observed spectrum minus calculated one) displayed on the right panel.

Figure \ref{profils_T_May2013}(a) shows the retrieved thermal profiles around the south pole in May 2013, 
with regions of maximum information displayed as solid lines. 
The 10 mbar level ($\sim$ 100 km) does not show strong meridional variations with temperatures in the 125 - 135 K 
range from 87$^{\circ}$S to 54$^{\circ}$S, while at 1 mbar ($\sim$170 km) we observe much larger meridional variations with a
temperature increase from 117 K at latitudes higher than 80$^{\circ}$S to 158 K at 54$^{\circ}$S. 
We derive a thermal inversion 
in the 10-0.5 mbar region at latitudes higher than 80$^{\circ}$S that is not observed for higher southern latitudes. A comparable thermal inversion 
was observed during the northern winter at high northern latitudes with a $\sim$1 km vertical resolution from 
Cassini radio occultation data \citep{Schinder_2012}.

Figure \ref{profils_T_May2013}(b) displays the thermal profile at 79$^{\circ}$S derived in early 2015 combining 
limb observations of March 2015 probing the 0.03-0.002 mbar region and nadir observations of December 2014 
probing the 0.2-20 mbar region. At this latitude, temperatures derived deeper than the 
0.2-mbar ($\sim$215 km) level are colder in early 2015 than in May 2013. 
For instance, between May 2013 and early 2015, temperature decreased by 15 K at 0.5 mbar and by 10 K at 10 mbar ($\sim$100 km).

In  May 2013 and early 2015, in the 80$^{\circ}$S-90$^{\circ}$S region, we retrieve stratospheric temperatures of 115 K around 
0.5 mbar, 
which corresponds to the 
coldest temperature that has been observed for these pressure levels over the entire Cassini mission.

\section{Gas mixing ratio retrievals}  \label{section_gaz_retrieval}

	\subsection{Retrieval from nadir spectra acquired in May 2013}

\begin{figure}[p] 
	\centering

	 \includegraphics[scale=0.7]{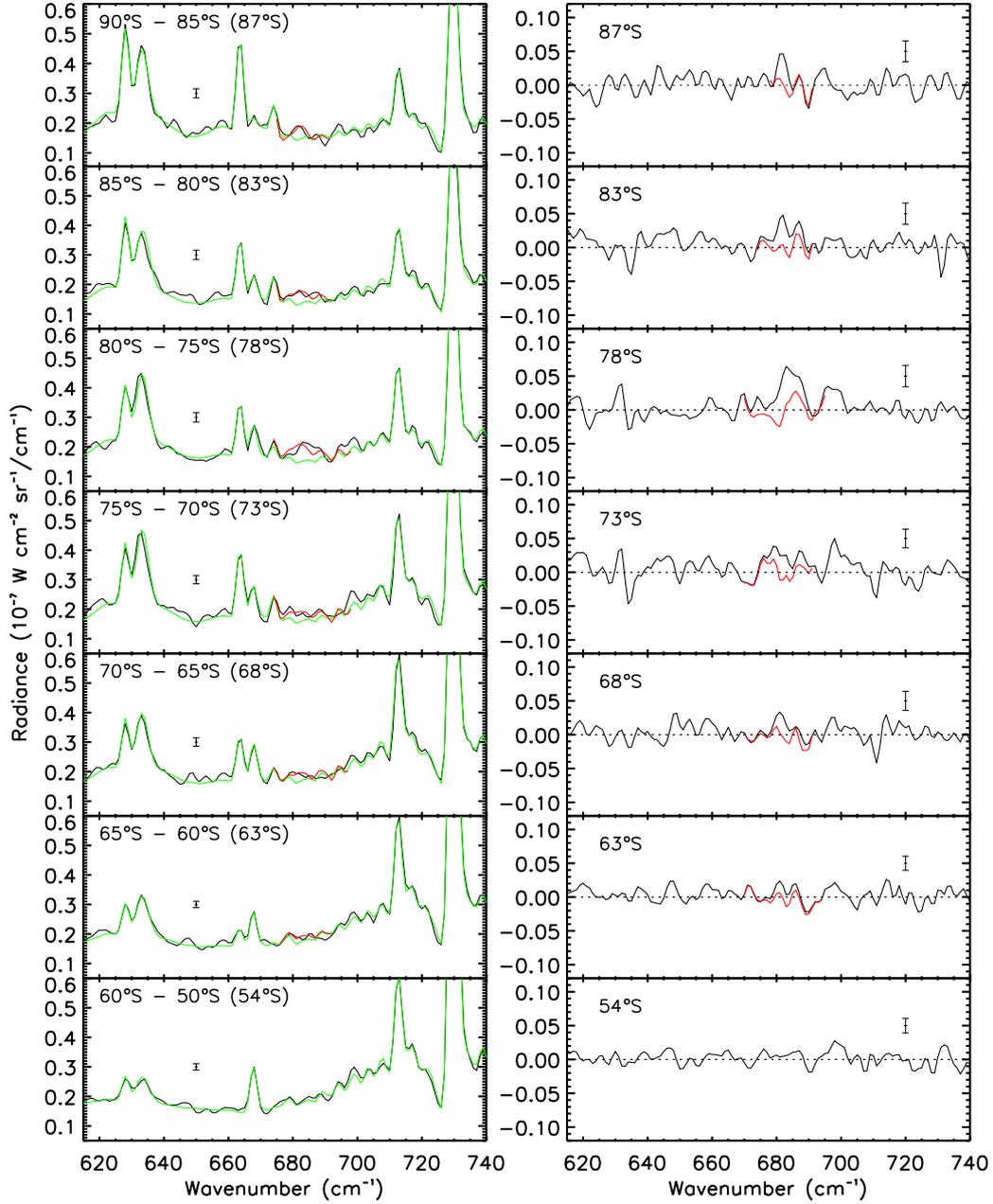}
	 \caption{CIRS nadir spectra in May 2013 (black line) of the 615-740 cm$^{-1}$ region compared to the calculated spectra without C$_6$H$_6$ ice (green) 
	 from 87$^\circ$S to 54$^\circ$S and including the C$_6$H$_6$ ice signature (red). Latitude ranges and 
	 mean latitude in brackets are given for each observed spectrum. Right: Corresponding residuals (observed spectrum minus calculated spectrum) 
	 without C$_6$H$_6$ ice (black) and including C$_6$H$_6$ ice (red). 
         1-$\sigma$ error bars are given for each spectrum.
		 }	
	\label{spe_FP3}
	\end{figure}

\begin{figure}[p] 
	\centering

         \includegraphics[scale=0.55]{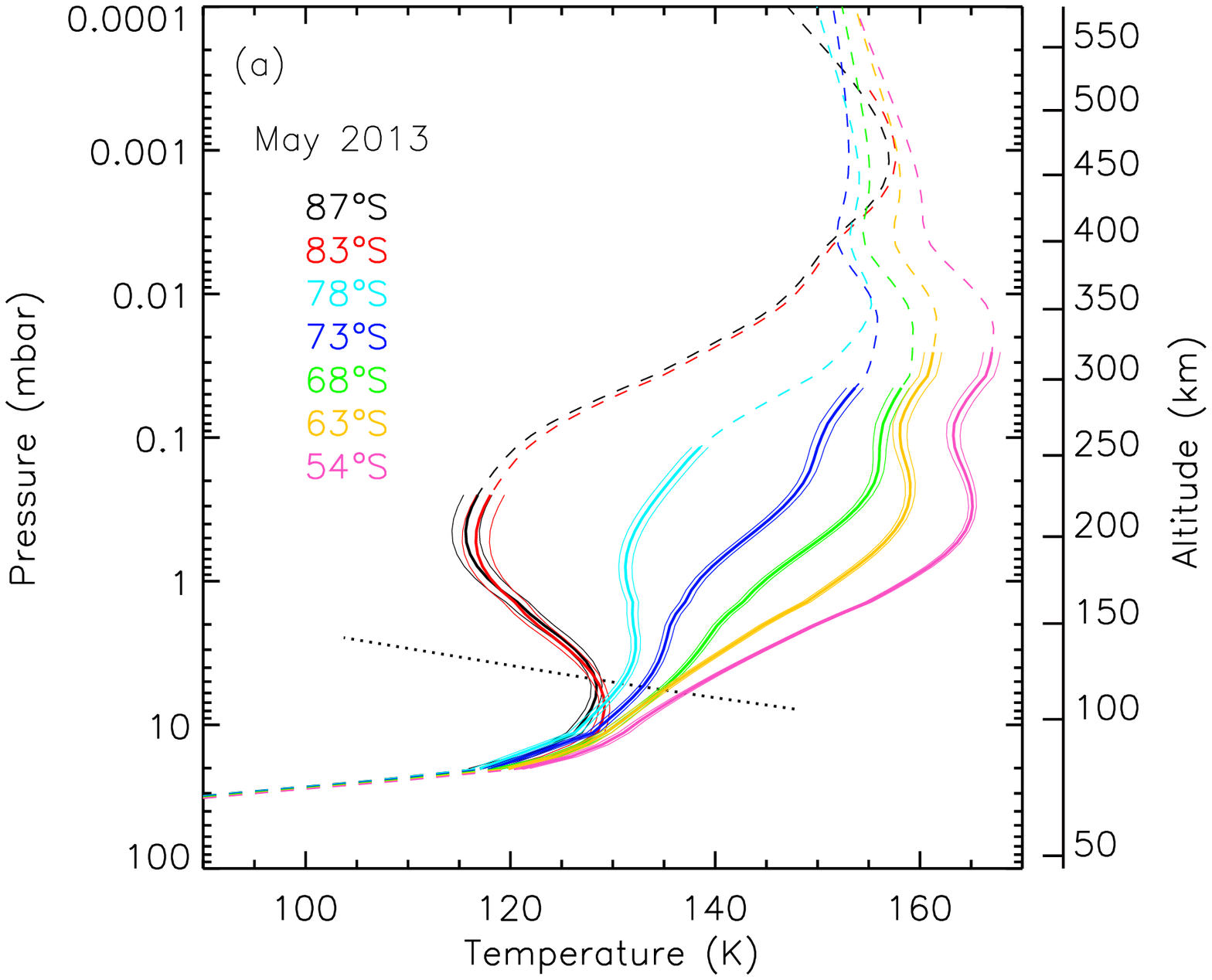}
        \includegraphics[scale=0.55]{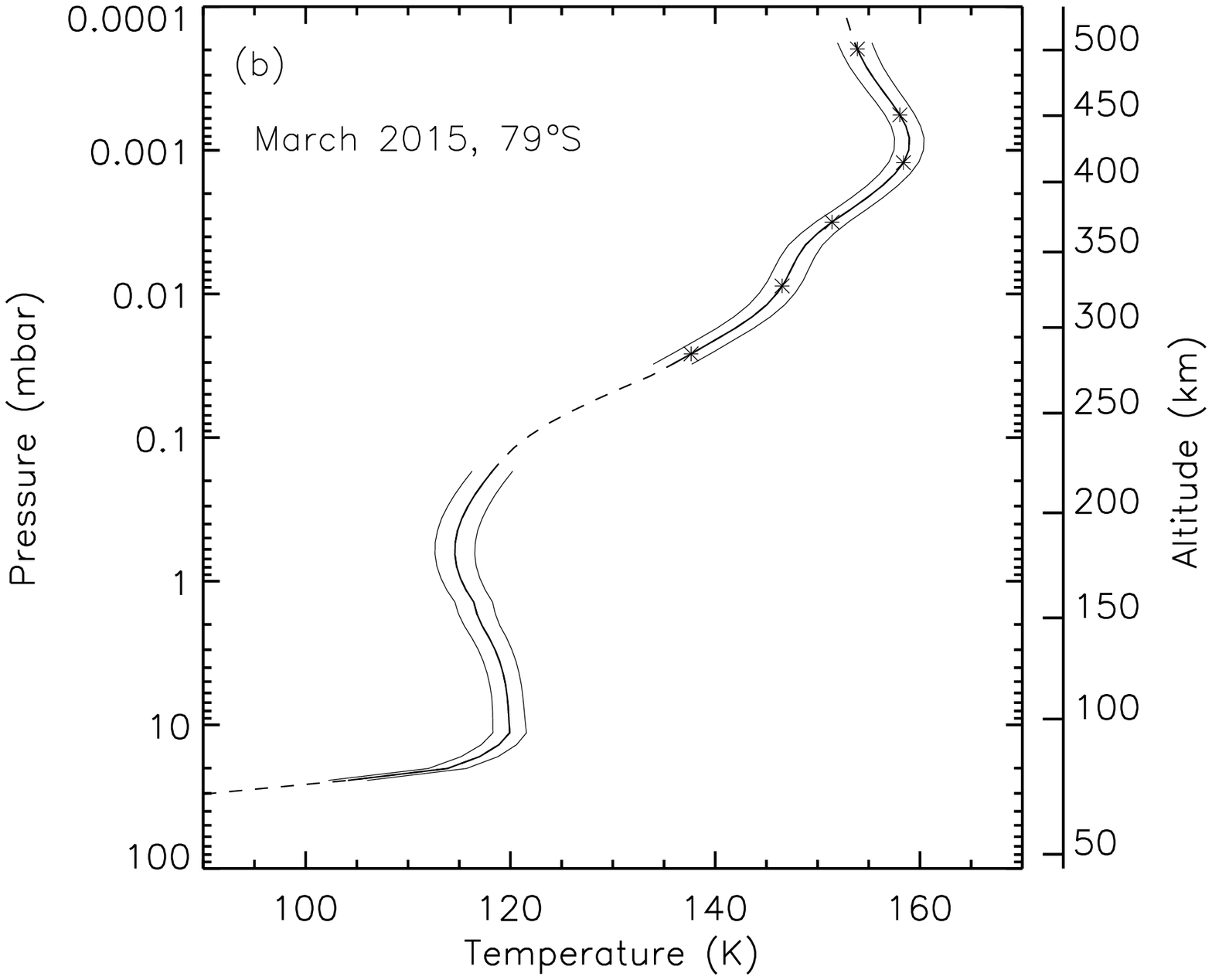}
	
	 \caption{(a): Thermal profiles around the south pole in May 2013. Altitude scale refers to the 78$^\circ$S
	 latitude. Thermal profiles relax to the a priori ones in regions without information (dashed lines). Dotted line
	 represents the dry adiabatic lapse rate.
	 (b) Thermal profile at 79$^\circ$S derived from 6 limb spectra, with tangent heights symbolized as crosses, 
	 and one nadir spectrum acquired in December 2014 at the same latitude, probing the pressure levels between 20 and 0.1 
	 mbar. Dashed lines give regions without information from CIRS data. 
		 }
	\label{profils_T_May2013}
	\end{figure}

The retrieved temperature profiles displayed in Figure \ref{profils_T_May2013}(a) were used to reproduce the thermal emission 
of gas ro-vibrational bands observed in spectra acquired by the CIRS FP3 focal plane.
Gas emission bands of C$_2$H$_2$ (730 cm$^{-1}$), C$_4$H$_2$ (628 cm$^{-1}$), CH$_3$C$_2$H (633 cm$^{-1}$), C$_6$H$_6$ (674 cm$^{-1})$, 
HCN (713 cm$^{-1}$), HC$_3$N (663 cm$^{-1}$) and CO$_2$ (668 cm$^{-1}$) are visible on the observed spectra 
displayed in Figure \ref{spe_FP3} (left). An inversion algorithm described by \cite{Vinatier_2007a, Vinatier_2015} was 
used to retrieve molecular volume mixing ratios from the fits of molecular emission bands and haze optical depth from the fit 
of the continuum. The haze optical depth was retrieved from the
600-620 cm$^{-1}$ and 640-660 cm$^{-1}$ spectral regions, using the spectral dependance of \cite{Vinatier_2012}.
Among molecules listed above, C$_2$H$_2$ is the only one for which we can derive vertical information from nadir spectra 
as it displays well defined P,Q and R-branches, with different opacities probing different pressure regions. For the other molecules, 
which display a single Q branch, it is not possible to constrain a vertical distribution. 
The choice of the a priori mixing ratio profiles therefore has an impact on the 
retrieved mixing ratios.  
Since June 2011, strong temporal variations were observed above the South pole. Strong enhancements of molecular gas volume mixing ratios 
were first observed in the 450-500 km region, and then gradually 
moved towards deeper levels, due to transport by the descending branch of the global circulation cell 
and confinement of the polar vortex \citep{Teanby_2012, Vinatier_2015, Achterberg_2014}.
For inversions of nadir spectra at latitudes higher than 80$^{\circ}$S, we used, as 
a priori vertical profiles, those that we retrieved from limb spectra acquired in September 2011 at 85$^{\circ}$S
 \citep{Vinatier_2015} and in March 2015 at 79$^{\circ}$S in the present study. For latitudes lower than 65$^{\circ}$S, we used as 
 a priori, the profiles that we derived in February 2012 at 46$^{\circ}$S from limb spectra as they are more representative 
 of the vertical profiles outside the polar vortex.
 Best fits of the averaged nadir 
spectra are displayed in Figure \ref{spe_FP3} (left panel) with their corresponding residuals 
(observed radiance minus calculated one) on the right panel. 

	\subsection{Spatial distribution of C$_6$H$_6$ gas mass mixing ratio at high southern latitude in May 2013}

The C$_6$H$_6$ gas mass mixing ratios derived for the 
seven latitudinal beams between 90$^{\circ}$S and 65$^{\circ}$S are displayed in Figure \ref{mixing_ratios_nadir} \textcolor{black}{(black line)
as well as their 1-$\sigma$ enveloppes (dotted lines) that include spectral noise contribution and uncertainty on temperature}. 
Benzene gas emission band at 674 cm$^{-1}$ is not detected in the 65-50$^{\circ}$S region (see Figure \ref{spe_FP3}). 
At levels deeper than the saturation level (level where the mixing ratio profile intercepts the saturation curve \textcolor{black}{displayed 
as a pink dashed line)}, benzene mass mixing ratio profiles follow the saturation law of \cite{Fray_Schmitt_2009}, 
calculated using the temperature profiles of Figure 2 (a). 
In the 90 - 80 $^{\circ}$S region, benzene saturation occurs at very unusually low pressure levels around 0.03 mbar 
(280 km), while for lower latitudes, it is observed deeper and deeper with saturation levels located at 0.2 mbar ($\sim$230 km) at 78$^{\circ}$S, 1 mbar (170 km) at 73$^{\circ}$S 
and 5 mbar (120 km) at 68$^{\circ}$S. 
These saturation levels are unusually high as benzene is predicted to condense around 80 km at mid-latitudes \citep{Barth_2017}.

       \subsection{Retrieval from limb spectra acquired in March 2015 at 79$^{\circ}$S} \label{retrieval_from_limb}
       
We used the retrieved temperature profile displayed in Figure \ref{profils_T_May2013}(b) to retrieve the 
vertically resolved profiles of molecular gas mixing ratios from limb spectra acquired in March 2015. We applied a shift 
of +40 km on the limb spectra line-of-sight altitudes extracted from the CIRS database, as derived from the best fit of 
the CH$_4$ band in the thermal profile retrieval step (see Section 3.1). 

We infer molecular mixing ratios highly enhanced for all molecules \citep{Vinatier_2016}, typically by a factor 5-10 
compared to what we derived near the south pole in September 2011 and February 2012 \citep{Vinatier_2015}. 
This is in agreement with results obtained by \cite{Teanby_2017}. 
We only present here results regarding benzene, while we will present mixing ratios profiles of other molecules in a 
future paper focusing on the seasonal variations of all molecular mixing ratios during the northern spring.  

Figure \ref{spe_FP3_limb_haut} displays the fit of the limb spectra acquired at high altitude in March 2015, where the C$_6$H$_6$
gas emission band is optically thin. 

       \subsection{Benzene gaz mass mixing ratio vertical profile in March 2015 at 79$^{\circ}$S} \label{C6H6_gaz_limb}

Figure \ref{mixing_ratios_limb} displays the retrieved C$_6$H$_6$ gas mass mixing ratio profile in the upper stratosphere and 
mesosphere in March 2015 at 79$^{\circ}$S. 
Molecular mixing ratios are so large in March 2015 that only the highest limb spectra, acquired above the 0.003 mbar level
($\sim$360 km), are optically thin. 
Deeper limb spectra have an optically thick contribution that probes in front of the limb tangent point along the 
line-of-sight, which corresponds to thermal emission coming from lower latitudes. We therefore plotted the C$_6$H$_6$ gas mass mixing ratio in the (0.1-3)$\times$10$^{-3}$ mbar region.

We derived a gas mass mixing ratio that increases from 3$\times$10$^{-6}$ at 0.003 mbar ($\sim$360 km) to 1$\times$10$^{-5}$ at 
3$\times$10$^{-4}$ mbar ($\sim$ 460 km). The C$_6$H$_6$ gas mass mixing ratio observed at 0.003 mbar is
comparable to the values derived from nadir observations at 83$^{\circ}$S and 87$^{\circ}$S in May 2013. 
%and seems to be larger than the 
%value derived from nadir observation at similar latitude (78$^{\circ}$S). This suggests an horizontal 
%extension of the polar vortex between 2013 and 2015, which is in agreement with observations of \cite{Teanby_2017} and \cite{Vinatier_2016}.

\section{Benzene ice spectral signature}

	\subsection{Detection}

As seen from fits and residuals of our fits of nadir spectra of May 2013 displayed in Figure \ref{spe_FP3}, 
calculated spectra (displayed in green) including gas emissions of C$_4$H$_2$, CH$_3$C$_2$H, HC$_3$N, CO$_2$, C$_6$H$_6$, HCN and C$_2$H$_2$ 
do not reproduce the 675-685 cm$^{-1}$
and 692-702 cm$^{-1}$ spectral ranges, with a larger misfit for higher latitudes for the 675-685 cm$^{-1}$ feature.
We believe that the 692-702 cm$^{-1}$ spectral signature is similar to the 
one observed by Voyager at 70$^{\circ}$N \citep{Coustenis_1999}. We also observed this signature with CIRS at the north pole
during the northern winter. Its origin is currently unknown and we will not consider it further in our 
analysis.
Our study presents the first observation of the 682 cm$^{-1}$ spectral signature. 
The only gas expected in Titan's atmosphere to display a spectral signature at this wavenumber is C$_2$H$_3$CN but its 
Q-branch is too narrow ($\sim$2 cm$^{-1}$) to reproduce the 682 cm$^{-1}$ spectral signature.
The fact that it is only observed at the latitudes of coldest temperatures
and has a relatively large width of $\sim$10 cm$^{-1}$ supports a vibrational mode of an ice as a spectroscopic candidate.
Indeed, 682 cm$^{-1}$ is the wavenumber of a vibrational mode of the C$_6$H$_6$ ice. 
%C$_6$H$_6$  gas had never been so enriched in Titan's stratosphere. 
At the south pole, C$_6$H$_6$ gas volume mixing ratio is enhanced by several orders of magnitude compared to 
lower latitudes where we usually derive upper limits of 0.3 ppb in the 2-5 mbar region \citep{Vinatier_2015}.
In the equatorial region, C$_6$H$_6$ gas condenses around 80 km \citep{Barth_2017}, while near the South pole, the combination 
of low temperature and high C$_6$H$_6$ gas volume mixing ratio makes this gas saturating around 0.03 mbar ($\sim$280 km)
at latitudes higher than 80$^{\circ}$S (see Figure \ref{mixing_ratios_nadir}).
Additionally, the 682 cm$^{-1}$ spectral signature is only seen at latitudes where the C$_6$H$_6$ gas emission is observed.
All of these arguments reinforce the interpretation of the 682 cm$^{-1}$ feature as due to, at least partly, C$_6$H$_6$ ice.

	\subsection{Model of the C$_6$H$_6$ ice spectral signature}
	
%There is currently no published optical constants of C$_6$H$_6$ ice in the mid-infrared. 
%We measured a preliminary spectral dependence of the 
%C$_6$H$_6$ ice imaginary refractive index from its laboratory measurements performed at 130 K.  
Thin films of solid pure C$_6$H$_6$ were condensed at 130 K (to insure crystallinity of the sample) from pure C$_6$H$_6$ gas in the cryogenic cell fitted in the FTIR spectrometer located at Institut de Plan\'etologie et d'Astrophysique de Grenoble, France \citep{Quirico_Schmitt_1997}. The film thickness was monitored by He-Ne laser interference.  Transmission spectra over the mid-IR (400-4500 cm$^{-1}$) have been recorded at 1 cm$^{-1}$ resolution between 130 and 20 K. 
Four main groups of bands were observed with one strong band around 680 cm$^{-1}$ attributed to the $\nu_4$ CH bending mode \citep{Bertie_2004}. 
This band is asymmetric with a width of about 6.5 cm$^{-1}$ and its peak has a blended double structure at 679 and 681 cm$^{-1}$. Their positions are only very weakly sensitive with temperature (shift $<$ 0.5 cm$^{-1}$ between 60 and 130 K). A weaker band (integrated intensity 20 times less) at 705.5 cm$^{-1}$ and another pair of bands at 1032 and 1038.5 cm$^{-1}$ ($\nu_{14}$ CH bending modes) are present in the CIRS spectral range.

In order to derive the spectral dependence of the real refractive index of C$_6$H$_6$ ice, we applied a substractive Kramers-Kronig
algorithm constrained with a visible real refactive index n$_r$ = 1.54 at 632.8 nm, derived by \cite{Romanescu_2010}.
Real and imaginary indices \citep{Schmitt_C6H6_data_2017} are available in the GhoSST (Grenoble Astrophysics and Planetology Solid Spectroscopy and Thermodynamics) database of the SSHADE
(Solid Spectroscopy Hosting Architecture of Databases and Expertise) database infrastructure\footnote{https://www.sshade.eu/; 

C$_6$H$_6$ ice spectrum:
https:/doi.org/10.17178/SSHADE.GHOSST.EXPERIMENT{\_}BS{\_}20170830{\_}001.V1}. 

Figure \ref{cross_sections} displays the spectral dependences of pure C$_6$H$_6$ ice extinction (top), absorption (middle) and scattering (bottom) cross sections 
per unit particle volume for spherical particles with different radii from 0.1 to 4 $\mu$m. 

For ice particles with radii smaller than 
1.0 $\mu$m, scattering is negligible and the extinction per unit particle volume does not depend on the particule
radius.
For radii larger than 1 $\mu$m, scattering contribution is not negligible, which results in a change
of the band shape more pronounced for particles with radii equal or larger than 1.5 $\mu$m, resulting in poor fit of the 
682 cm$^{-1}$ emission band.  
We choose here to model the C$_6$H$_6$ ice grains by spheres of 0.5-$\mu$m radius.

\section{Benzene ice mass mixing ratios}   \label{resultats_profils_C6H6}

\subsection{Retrieval method}

Our code retrieves ice optical depth in each layer of the pressure grid from the best fit of the 682 cm$^{-1}$ spectral 
signature. Retrievals were performed for both nadir and limb spectra assuming a spherical shape with a 0.5-$\mu$m radius for the 
C$_6$H$_6$ ice particles. As C$_6$H$_6$ ice particles appear when the C$_6$H$_6$ gas mixing ratio becomes saturated, or in other words 
where the mixing ratio profiles displayed in Figure \ref{mixing_ratios_nadir} intercept the saturation 
curve, we applied a cutoff in the a priori vertical profile of the C$_6$H$_6$ cloud optical depth at the level where C$_6$H$_6$ becomes saturated. 
Above this level, we assumed a zero optical depth. For the May 2013 nadir observation, this level varies from  
0.03 mbar (280 km) at 87$^{\circ}$S and 83$^{\circ}$S to 5 mbar (120 km) at 68$^{\circ}$S, while for March 2015 limb observations
we applied a cutoff at 0.02 mbar.

Below the saturation level, we do not know the cloud vertical distribution and we cannot derive any vertical profile 
from nadir observations of the C$_6$H$_6$ ice spectral signature. Several slopes of the vertical distribution 
can yield similar satisfactory fits of a nadir spectrum.  
We therefore first derived constraints on the vertical distribution of the cloud optical depth from the limb 
observations acquired in March 2015, where the 682 cm$^{-1}$ spectral band is observed in limb spectra at 278 km and deeper
(see Figure \ref{spe_FP3_limb}).   

But, as mentioned in Section \ref{retrieval_from_limb}, emission bands of molecular gases, which are highly enriched, 
generally become opaque deeper than the 0.003 mbar level ($\sim$360 km), like for C$_6$H$_6$ gas (see Section 4.4). 
It would be possible to derive information at deeper levels by using a 2-D 
retrieval algorithm, as performed by \cite{Achterberg_2008}, which would take into account the thermal and compositional meridional gradients along
the limb line-of-sight. Building such a 2-D retrieval algorithm is outside the scope of this paper. Nevertheless, C$_6$H$_6$ ice band and  
continuum emission due to haze are optically thin at levels in the 0.03 - 0.6 mbar (270 - 175 km) region, much deeper 
than the region were gas emissions are optically thin. As we are interested in retrieving the vertical distribution of
the C$_6$H$_6$ ice cloud, we chose to retrieve gas mixing ratios using limb spectra deeper than 0.003 mbar individually but knowing that
the retrieved gas mixing ratios were not representative of the 79$^{\circ}$S latitude, but instead lower latitudes.
The idea here is to obtain the best fit of each limb spectrum acquired below the 0.03 mbar level in order to fit the 682 cm$^{-1}$ ice signature in a second step.
Best fits of limb spectra acquired between 168 and 320 km are displayed in Figure \ref{spe_FP3_limb}. These limb spectra
were initially acquired at 0.5 cm$^{-1}$ resolution and degraded to the spectral resolution of 2.8 cm$^{-1}$ in order to 
facilitate comparison by eye with the nadir spectra of Figure \ref{spe_FP3}. 

After having obtained the gas mass mixing ratio profiles that best fit the 168-320 km limb spectra in March 2015, we derived the 
vertical distribution of the C$_6$H$_6$ ice cloud by fitting the 678 - 683 cm$^{-1}$ region. This vertical ice cloud profile was then 
used as an a priori for the retrievals of the ice mixing ratio spatial distribution from nadir spectra acquired in May 2013.

\subsection{Results}  \label{section_results}

Figures \ref{spe_FP3} and \ref{spe_FP3_limb} display fits of the observed 682 cm$^{-1}$ 
spectral signature in May 2013 in nadir spectra and in March 2015 in limb spectra.  
Residuals are displayed in the right
panel of both figures without the benzene ice contribution (black line) and including it (red).  
In May 2013 (Figure \ref{spe_FP3}), the 682 cm$^{-1}$ ice feature as well as the unidentified 695 cm$^{-1}$ one are more 
prominent in the 80-75$^{\circ}$S latitudinal region. The 682 cm$^{-1}$ C$_6$H$_6$ ice spectral signature is not detectable for latitudes lower than 
65$^{\circ}$S, where the benzene gas emission at 674 cm$^{-1}$ also disappears.
In March 2015 (Figure \ref{spe_FP3_limb}), at 79$^{\circ}$S, the 682 cm$^{-1}$ spectral signature is mixed with several 
unidentified spectral contributions in limb spectra at 168, 205 and 239 km with 
apparently two new spectral signatures at $\sim$687 cm $^{-1}$ and $\sim$702 cm$^{-1}$. The 695 cm$^{-1}$
signature, which is observed at the same latitude in May 2013 in nadir spectra, also possibly contributes to the observed emission.

\begin{figure}[p] 
	\centering

         \includegraphics[scale=0.74]{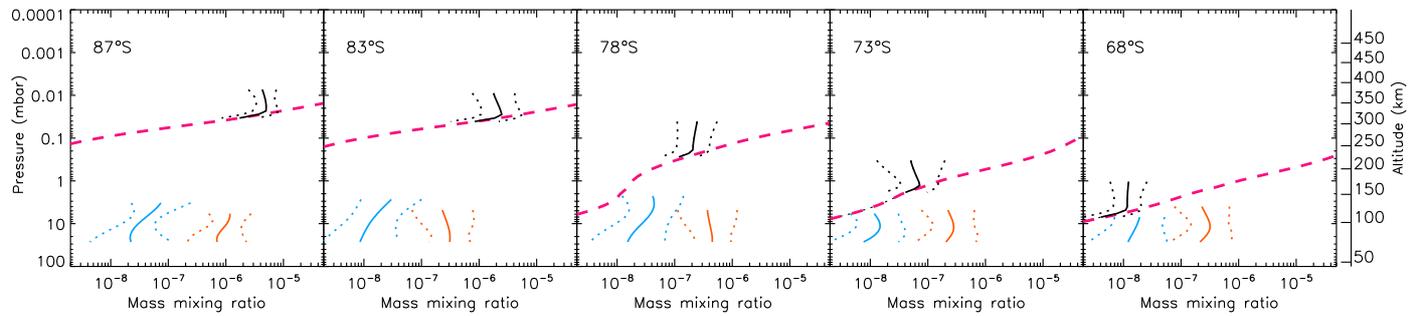}
    
	 \caption{Retrieved mass mixing ratio of C$_6$H$_6$ gas (black)  
	 with the saturation curves for each latitude displayed in pink dashed lines. 
	 The retrieved haze and C$_6$H$_6$ ice mass mixing ratios are 
	 displayed in orange and blue, respectively. Solid lines give the mean mass mixing ratio, while the dotted lines give the 
	 1-$\sigma$ error bars. 
	 %Dashed lines for 87$^\circ$S and 83$^\circ$S give the profile for pressure level probed by limb spectra.  
	 The vertical altitude scale corresponds to the 68$^\circ$S latitude.}	
	\label{mixing_ratios_nadir}
	\end{figure}

\begin{figure}[p] 
	\centering
	 
	 \includegraphics[scale=0.9]{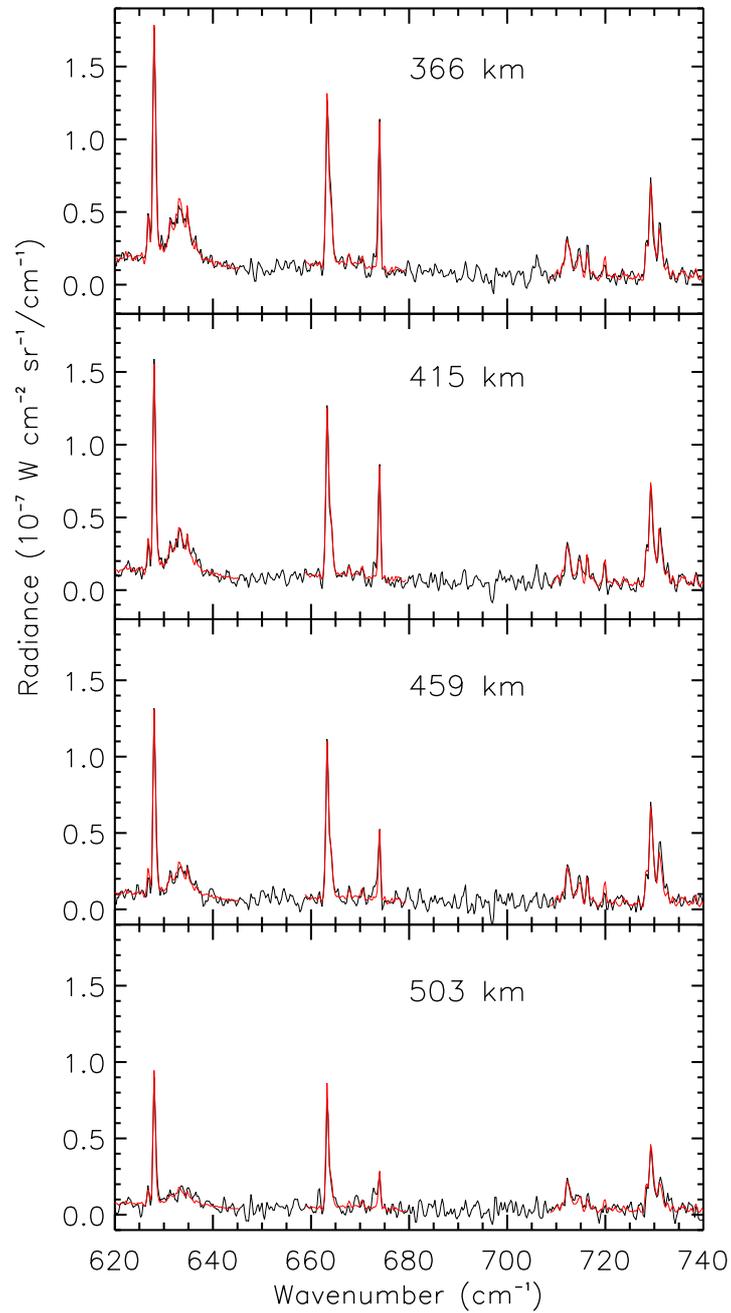}
	 \caption{CIRS limb spectra at 79$^\circ$S in March 2015 (black line) of the 615-740 cm$^{-1}$ region compared 
	 to the calculated spectra (red). Altitudes of the averaged spectra
	 take into account the +40 km shift related to the altitude given in the CIRS database. 
		 }	
	\label{spe_FP3_limb_haut}
	\end{figure}

\begin{figure}[p] 
	\centering

         \includegraphics[scale=0.6]{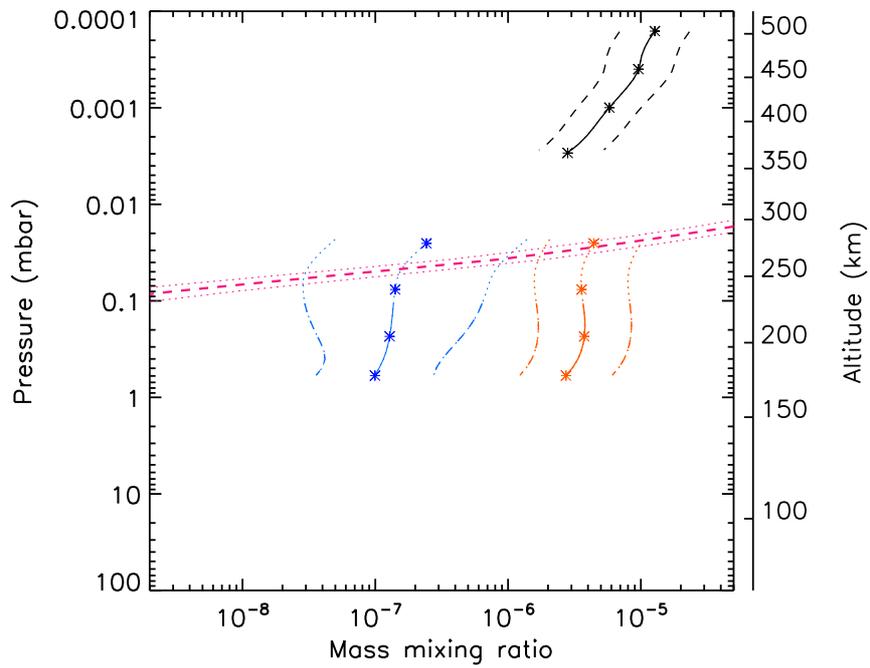}
    
	 \caption{Retrieved C$_6$H$_6$ gas mass mixing ratio profile (black line) from the 79$^{\circ}$S in March 2015 limb spectra 
	 of Figure \ref{spe_FP3_limb_haut}
	 with the calculated saturated mass mixing ratio in pink dashed lines. The retrieved haze and C$_6$H$_6$ ice mass mixing ratios profiles are 
	 displayed in orange and blue, respectively. Altitudes of the line-of-sights of the mean limb spectra are displayed as crosses.
	 Solid lines give the mean mass mixing ratios, while dotted lines give the 
	 1-$\sigma$ error bars.}	
	\label{mixing_ratios_limb}
	\end{figure}

We derive the C$_6$H$_6$ ice mass mixing ratio profile from the retrieved optical depth d$\tau$ in 
each layer of altitude thickness dz from the best fit of the observed spectra. 
In each layer, the number density of ice particles is equal to d$\tau$/($\sigma_{682}$ dz), where 
$\sigma_{682}$ is the extinction cross section of the ice particle at 682 cm$^{-1}$ for a spherical particle of 0.5 $\mu$m
radius (see Figure \ref{cross_sections}). 
In May 2013, from the nadir data between 87$^{\circ}$S and 68$^{\circ}$S, at 7 mbar we derive ice number densities of 0.4 to 1.3 
particles/cm$^{3}$, about 10 times smaller than the haze number density at the same level. In March 2015, from the limb observation, we derive at 0.7 mbar an ice particle 
number density of 0.4 cm$^{-3}$, also 10 times lower than the haze number density there. 
The mass of an ice particle is derived assuming a C$_6$H$_6$ ice density 
of 1.1 g.cm$^{-3}$, which lies between the values 1.094 and 1.114 g.cm$^{-3}$ measured at 138 K and 77 K, 
respectively \citep{Cox_1958, Bacon_1964, Romanescu_2010}. The retrieved C$_6$H$_6$ ice mass mixing ratio profiles derived in May 2013 and March 2015 are displayed 
in Figures \ref{mixing_ratios_nadir} and \ref{mixing_ratios_limb} (in blue), respectively.
\textcolor{black}{1-$\sigma$ error envelopes on profiles derived from nadir data include contributions from noise and temperature uncertainty and those derived from limb observations additionally incorporate contribution of the vertical shift uncertainty (see Section 3.1)}. 

Limb spectra in March 2015 probe the 0.02 - 5 mbar range, with lesser 
constraints in the 0.02 - 0.07 mbar range because of the poorly constrained temperature profile in this range (see Figure \ref {profils_T_May2013} (b)). We can notice that the 278 km limb spectrum probes a level located close to the saturation curve (see Figure \ref{mixing_ratios_limb}), 
while the 
320-km limb spectrum probes more than one scale height above the saturation curve, which is compatible with the 
fact that we do not observe the C$_6$H$_6$ ice spectral feature in this limb spectrum. In the 0.1-1 mbar region,
we derive a roughly constant C$_6$H$_6$ ice mass mixing ratio of $\sim$1$\times$10$^{-7}$.

In May 2013, the derived C$_6$H$_6$ ice mass mixing ratio at 10 mbar ($\sim$100 km), 
which is the level where the C$_6$H$_6$ ice thermal emission mostly comes from, seems to be constant within error bars from the South pole to 65$^{\circ}$S and equal to 
$\sim$1-2$\times$10$^{-8}$ (see Figre 4).

Figures \ref{mixing_ratios_nadir} and \ref{mixing_ratios_limb} additionally display 
the haze mass mixing ratio profiles (orange lines) derived at the same pressure levels as the ice cloud. 
Calculation of the aerosol mass mixing ratio assumes particles of 3000 monomers of 0.05 $\mu$m
radius each and a material density of 0.6 g cm$^{-3}$.
For all latitudes observed in May 2013 and also in March 2015 at higher altitude, 
we derive that haze mass mixing ratio is always higher than the C$_6$H$_6$ ice mass mixing ratio, with a typical factor of 20-30. 
These similar ratios from one latitude to another and at two different dates suggest homogeneous composition 
inside the south polar vortex.    
In March 2015, we derive quite constant-with-height C$_6$H$_6$ ice and haze mass mixing ratios profiles in the 0.1-0.7 mbar region, 
which is probably due to the air subsidence that bring enriched air in haze from the higher stratosphere to deeper levels and 
that aliment the C$_6$H$_6$ cloud with benzene gas enriched air from above.

\section{Discussion}

  \subsection{Benzene gas mixing ratio in the southern polar region in May 2013 and March 2015}
  
It is more appropriate to work with volume mixing ratio when dealing with the gas distribution, in which case the C$_6$H$_6$ 
gas mass mixing ratio profiles plotted in Figures \ref{mixing_ratios_nadir} and \ref{mixing_ratios_limb} 
 should be multiplied by a factor of 0.36
(ratio of the mean atmospheric molar mass equal to 27.79 g mol$^{-1}$, mostly due to N$_2$, over the benzene molar mass equal to 78 g mol$^{-1}$) 
to be converted into volume mixing ratio. 

In May 2013, at 87$^{\circ}$S the corresponding volume mixing ratio is $\sim$1.5$\times$10$^{-6}$ at 0.015 mbar ($\sim$ 300 km) and 
in March 2015, we derive an increase of the volume mixing ratio from 1$\times$10$^{-6}$ at 0.003 mbar ($\sim$370 km) to 
5$\times$10$^{-6}$ at 1.5$\times$10$^{-4}$ mbar ($\sim$507 km).

In March 2015, the C$_6$H$_6$ gas volume mixing ratios observed at 0.003 mbar is higher by a factor of 60 than the 
value observed in September 2011 at 85$^{\circ}$S at the same pressure level \citep{Vinatier_2015}. As mentioned earlier, no other limb 
data were acquired between these dates because of highly inclined orbits of Cassini. This increase is explained 
 by the subsidence bringing enriched air from high altitude toward deeper levels. 
 
 In May 2013 and March 2015, 
our derived C$_6$H$_6$ gas volume mixing ratios are comparable to the in situ values of 9$\times$10$^{-7}$ - 3$\times$10$^{-6}$ measured by the Cassini Ion and Neutral Mass Spectrometer at 1000 km \citep{Vuitton_2007, Cui_2009, Magee_2009}. 
Benzene is not the only molecule to display such a high enhancement inside the polar vortex: 
C$_2$H$_2$, HCN, HC$_3$N, C$_4$H$_2$, CH$_3$CCH, C$_2$H$_4$ are also observed to be highly enriched in March 2015 above 400 km \citep{Vinatier_2016}, with mixing ratios comparable to the INMS measurements around 1000 km. 
The highly enriched volume mixing ratios observed with CIRS inside the southern polar vortex, suggest that
the vortex barrier is very efficient not only at stratospheric levels but also at much higher altitude, so that the air observed in the 
stratosphere around 350 km has a similar composition as the air around 1000 km.

  \subsection{Benzene condensation level and the altitude of the south polar cloud}

In May 2013, from the nadir observations we can infer information on the altitude of the top of the cloud from the fit 
of the C$_6$H$_6$ gas emission band and the calculated saturation curve (see Section 4.1 and 4.2). In order to test the impact of the 
a priori C$_6$H$_6$ gas mixing ratio profile on the altitude of the condensation level, we tested two types of
a priori profiles: profile derived from limb spectra acquired in March 2015 (Figure \ref{mixing_ratios_limb}) 
extrapolated for deeper levels than the 0.003-mbar level and constant-with-height mixing ratio profiles. 
For both a priori profiles and at all latitudes, we derived similar condensation levels (within error bars) and similar volume 
mixing ratios at these pressure levels. Therefore, the tendency to observe condensation of C$_6$H$_6$ deeper and deeper 
while moving away from the South pole seems to be robust and is due to warmer temperature in the stratosphere for lower latitude. Thus,
the C$_6$H$_6$ cloud top should be located deeper with increasing distance from the pole.

In March 2015, from the limb observations, we can estimate the altitude of the top of the C$_6$H$_6$ cloud from the
altitude of the highest limb line-of-sight where the 682 cm$^{-1}$ benzene ice signature is observed. From these observations,
the cloud top is located near 0.025 mbar (278 km), which is similar to the condensation levels observed in May 2013 at 
latitudes higher than 80$^{\circ}$S.

In May 2013, among all molecules cited above, C$_6$H$_6$ is the one that condenses at the highest altitude, with condensation occurring at 
$\sim$0.03 mbar ($\sim$280 km) for a saturated volume mixing ratio of $\sim$2$\times$10$^{-6}$
in the 90$^{\circ}$S - 80$^{\circ}$S region, which is compatible 
with the derived altitude of 300 km of the south polar cloud observed in mid-2012 by the Cassini ISS instrument from the south 
pole to 80$^{\circ}$S \citep{West_2016} and the HCN cloud observed by Cassini/VIMS \citep{de_Kok_2014}. 
From our CIRS nadir dataset of May 2013 at 87$^{\circ}$S, we derive a condensation level for HCN located near 0.1 mbar ($\sim$235 km), 45 km
deeper than the C$_6$H$_6$ saturation level, with a saturated volume mixing ratio of $\sim$2$\times$10$^{-6}$, which is comparable 
to the C$_6$H$_6$ one at $\sim$0.03 mbar. Therefore, C$_6$H$_6$ ice is probably an important contributor to the composition of the upper
levels of the southern fall stratospheric polar cloud.  

In March 2013 at 85$^{\circ}$S and 80$^{\circ}$S, we observe a thermal inversion in the deep stratosphere at $\sim$10 mbar, 
which could make the atmosphere unstable regarding convection. Additionally, \cite{West_2016} mentioned convective
patterns in the ISS images of the South polar cloud, which could be explained by such a thermal inversion.
Figure \ref{profils_T_May2013} (a) shows as a black dotted line the dry 
adiabatic lapse rate. Comparison of slopes of the retrieved thermal profiles at 87$^{\circ}$S and 83$^{\circ}$S
and the dry adiabat shows that no convection should occur in the cloud. Nevertheless, we recall here that 
the vertical resolution is very limited because of nadir geometry and it is not excluded that a steeper gradient takes place over a limited
vertical range, similar to the one observed from radio occultation measurements at 74$^{\circ}$N during the northern winter \citep{Schinder_2012}.

%	\subsection{Dependence of the C$_6$H$_6$ ice spectrum with particle shape and composition} \label{section_spectre_particl_shape}
        \subsection{Dependence of the C$_6$H$_6$ ice spectrum with particle size and shape} \label{section_spectre_particl_shape}

%\subsubsection {Particle size and shape}

\begin{figure}[p] 
	\centering
	 \includegraphics[scale=0.7]{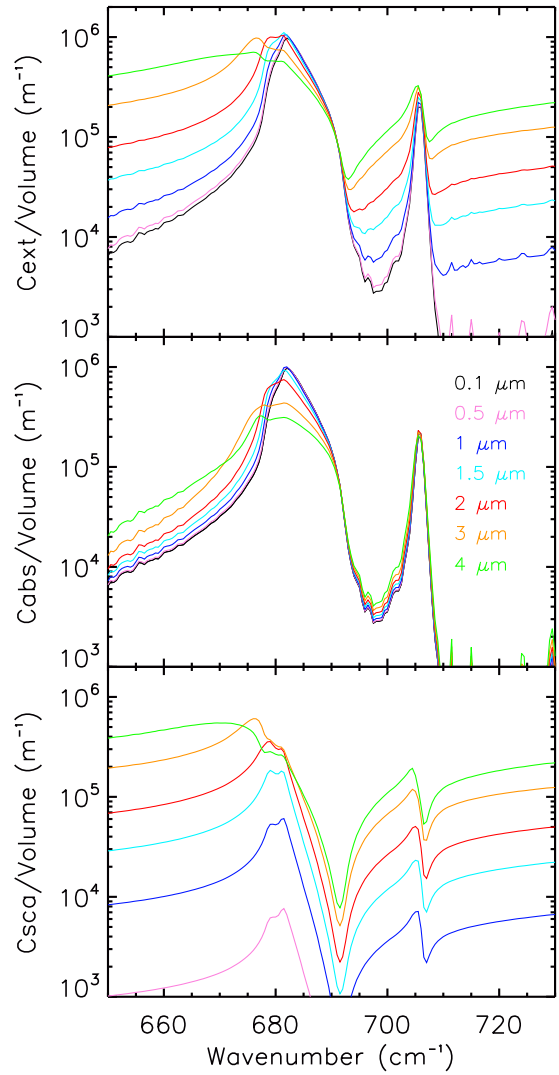}
         \caption{Spectral dependences of extinction (top), absorption (middle) and scattering (bottom) cross sections per unit particle volume calculated for spherical
	 particules composed of pure C$_6$H$_6$ ice with different radii from 0.1 $\mu$m (black) to 4 $\mu$m (green). 
			 }	
	\label{cross_sections}
	\end{figure}

\begin{figure}[p] 
	\centering
	 
	 \includegraphics[scale=0.7]{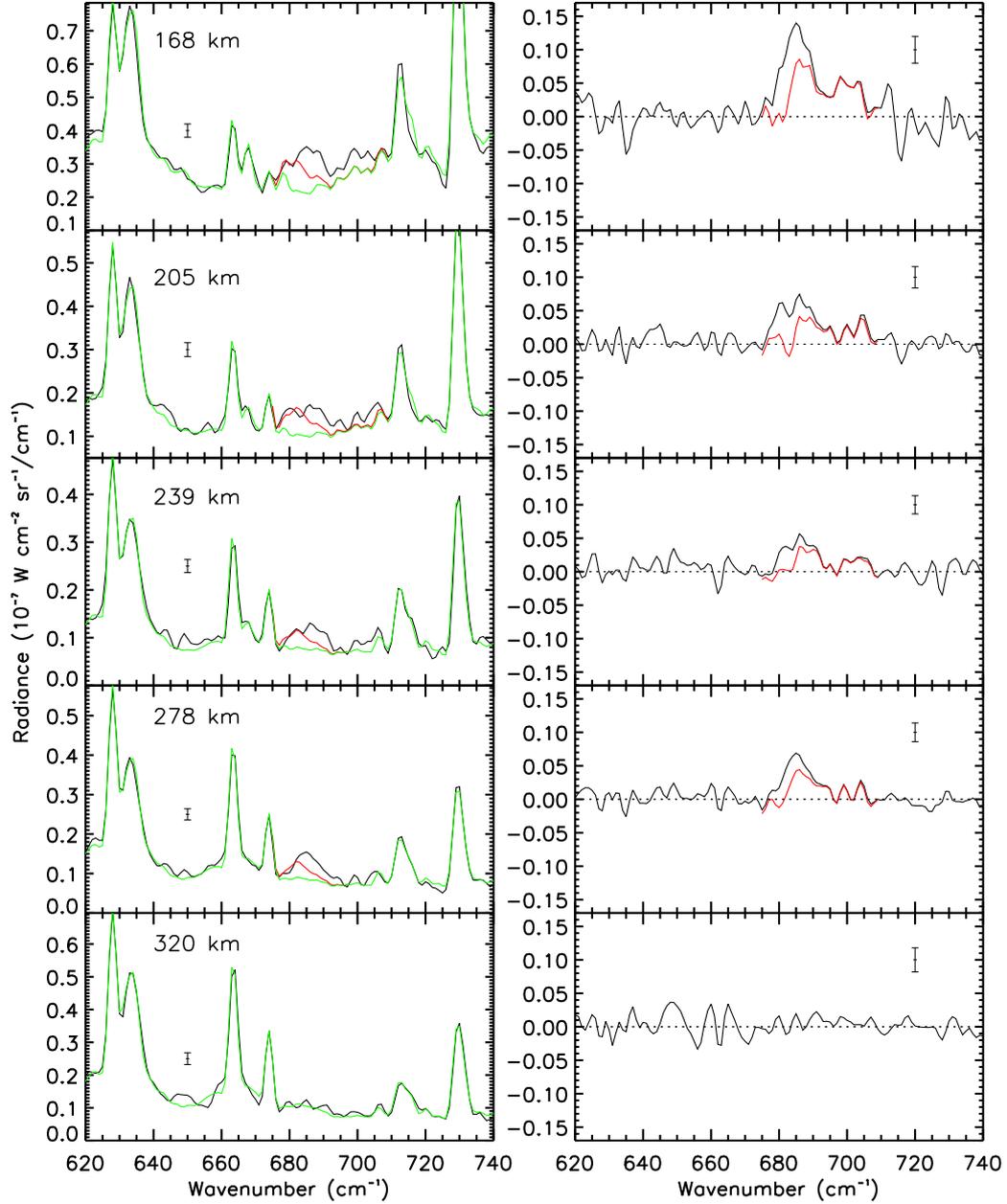}
	 \caption{CIRS limb spectra at 79$^\circ$S in March 2015 (black line) of the 615-740 cm$^{-1}$ region compared 
	 to the calculated spectra without C$_6$H$_6$ ice (green) and including it (red). Altitudes of the averaged spectra
	 take into account the +40 km shift relatively to the altitude given in the CIRS database. 
	 Right: Corresponding residuals (observed spectrum minus calculated spectrum) 
	 without C$_6$H$_6$ ice (black) and including it (red). 1-$\sigma$ error bars are given for each spectrum.
		 }	
	\label{spe_FP3_limb}
	\end{figure}

Figure \ref{cross_sections} shows the spectral dependence of calculated cross sections per unit particle volume for different radii of spherical particles made of pure 
C$_6$H$_6$ ice. For spherical particles of 0.1 and 0.5 $\mu$m radii, extinction cross sections per unit volume are very similar. 
This is expected as these radii are much smaller than the observed wavelength (682 cm$^{-1}$ corresponds to 
14.7 $\mu$m) and the imaginary part of the ice refractive index is not negligible (here equal to $\sim$1.5 for the maximum
of the absorbing band), a case in which the calculated extinction cross section does not depend on the shape 
of the ice particle, but only on its equivalent volume and can be described by Equation 8.4.2 of \cite{Hanel_2003_book}. 
Extinction is then dominated by absorption and scattering is negligible. 

For spherical particles with radii of 1.5 $\mu$m or smaller, we derived similar fits of the observed spectra and 
similar ice mass mixing ratios. For radii of 2 $\mu$m or larger, fits of the 682 cm$^{-1}$ 
spectral signature were degraded as the scattering contribution increases with increasing particle radius (Figure
\ref{cross_sections}), which results in a change in the shape of the spectral band, with almost no more extinction feature at 
682 cm$^{-1}$ for a 4 $\mu$m particle radius. In conclusion, we can set an upper limit of $\sim$1.5 $\mu$m for the equivalent 
radius of C$_6$H$_6$ ice particles from fits of CIRS spectra. 

In order to investigate the potential impact of the shape of the C$_6$H$_6$ ice particles, we additionally performed 
cross section calculations using DDSCAT 7.3 \citep{Draine_1994, Draine_2008, Flatau_2012} with cubic, rectangular, 
elliptical and with two-sphere or two-ellipse shape crystals made 
of pure C$_6$H$_6$ ice all having the same volume as a sphere of 0.5 $\mu$m radius.
% which is the equivalent size of a typical aerosol (3000 monomers each of 0.05 $\mu$m radius). 
The calculated extinction cross sections per unit particle volume  
slightly differ in shape from the spherical case, but with no possible discrimination from the observed 
CIRS emission.

  \subsection{Benzene ice particle size and sedimentation}

Satisfactory fits of CIRS spectra require C$_6$H$_6$ particles of equivalent radii smaller than 1.5 $\mu$m.
All results presented here were performed for C$_6$H$_6$ ice particle equivalent radii of 0.5 $\mu$m.
We can estimate theoretically the mean size of a C$_6$H$_6$ ice particle from the comparison of the condensation growth timescale 
($\tau_{cond}$, proportional to the particle radius), which is the timescale over which a particle of C$_6$H$_6$ ice increases its mass by a factor e, 
and the falling timescale ($\tau_{fall}$) that include the contribution of sedimentation timescale ($\tau_{sed}$, varying as the inverse of radius) 
over the saturated benzene gas scale height and the dynamical timescale ($\tau_{dyn}$) due to the air subsidence, with  $\frac{1}{\tau_{fall}}= \frac{1}{\tau_{sed}}+ \frac{1}{\tau_{dyn}}$, 
as the falling velocity is the sum of the sedimentation speed and the subsidence velocity.

\subsubsection{Dynamical timescale}

Vertical transport associated with the descending branch replenishes the C$_6$H$_6$ ice cloud from the top with air enriched in C$_6$H$_6$ gas. 
We can estimate the dynamical timescale of this transport over one scale height from the temporal evolution of the molecular 
gas mixing ratios profiles, as explained in \cite{Vinatier_2015}. We therefore need to use profiles derived with high
vertical resolution from limb observations. As before March 2015, no limb data of the South pole have been acquired
since September 2011, we preferred to use the closest limb observations of September 2015 \citep{Vinatier_2016} to estimate the 
downward velocity in the polar vortex in the March-September 2015 period. To derive this velocity, we used the derived gas mixing ratios 
of C$_6$H$_6$, HC$_3$N, C$_4$H$_2$, C$_3$H$_4$, HCN and C$_2$H$_2$ at the reference pressure of 0.008 mbar in March 2015 and 
infer the pressure levels at which these molecular mixing ratios were observed in September 2015. We then derived that 
the mean pressure level of the
same air composition was observed at a pressure level 4.5 times deeper in September 2015 than in March 2015. As the mean 
scale height is H$\sim$ 45 km at 0.02 mbar, we can derive the vertical distance over which the air moved downward 
equal to H$\times$ln(4.5)=1.5 H, 
corresponding to a downward vertical velocity of $\sim$4 mm s$^{-1}$ and $\tau_{dyn} \sim 10^{7}$ s at a pressure level of 0.02 mbar.
From this value, we can estimate that the vertical velocity at 0.1 mbar would be 0.8 mm s$^{-1}$ assuming that in the 0.1-0.01 mbar 
the flow convergence or divergence inside the polar vortex are small. This result is in agreement with the 
subsidence velocities that \cite{Teanby_2017} calculated in 2015 from energy balance considerations.
We assume here that $\tau_{dyn}$ in May 2013 was similar to the dynamical timescale derived in 2015.

\subsubsection{Sedimentation timescale}

We can estimate the sedimentation timescale, corresponding to the time needed 
for a particle to fall over one pressure scale height, for C$_6$H$_6$ ice particles.
We focus here on the 0.025 mbar pressure level ($\sim$285 km), which corresponds to the upper part of the C$_6$H$_6$ ice cloud
observed in May 2013 at 87$^{\circ}$S and 83$^{\circ}$S, from nadir spectra. At this low atmospheric pressure, 
we first check whether the interaction between an
ice particle and the ambient air should be treated as interaction with a continuum flow 
(particle radius $>>$ gas mean free path) or if the gas is too rarefied (particle radius $<<$ gas mean free path) in which case
we have to consider the gas kinetic theory.
In Titan's atmosphere, assumed here to be only made of N$_2$, at 0.025 mbar and 138 K, the 
mean free path of a molecule is $\sim$1.7 mm, which is larger by a factor of $\sim$1100 (equal by definition to the 
Knudsen number) than the maximum ice particle radius of 1.5 $\mu$m that was derived from our observations. 
%Therefore, an ice particle can be regarded from gas molecules 
%as another molecule and the Maxwell-Boltzmann statistics is needed to derive its sedimentation timescale. 
We are then here in the Knudsen regime and we therefore used Eq. (19) of \cite{Rossow_1978} to estimate this 
sedimentation timescale over one pressure scale height:

\begin{equation}
	\tau_{sed} = 27 \pi \rho_{atm} \bigg(\frac{2 k T}{\pi m}\bigg)^{\frac{3}{2}} \frac{1}{16 \rho_{ice} g^2 a} 
\end{equation} 

were $\rho_{atm}$ and $\rho_{ice}$ are the atmospheric and ice densities, respectively, $m$ is the mass of an 
atmospheric molecule, $g$ is the acceleration of gravity, $a$ is the particle radius and $T$ the 
atmospheric temperature.
We used $\rho_{ice}$ = 1100 kg m$^{-3}$, g = 1.1 m s$^{-2}$ determined at 
285 km (altitude of the 0.025 mbar level) and a temperature of 138 K and derive $\tau_{sed}$ = $\frac{10^{6}}{a}$ s, with $a$ in $\mu$m.
This corresponds to a sedimentation velocity H/$\tau_{sed}$, where H is the pressure scale height ($\sim$38 km at 0.025 mbar), 
of 1.9 cm s$^{-1}$ at 0.025 mbar.
$\tau_{sed}$, which is calculated for the idealized case of a sphere should be regarded as 
a lower limit of the sedimentation timescale as we have no idea of the shape of the C$_6$H$_6$ ice particles: for instance, they could be elongated or 
possibly condense on some parts of fractal aerosols that are quite fluffy and therefore would sediment less rapidly.

\subsubsection{Estimation of the particle size from falling and condensation timescales}

We estimated the condensation timescale using Eq. (17) of \citep{Rossow_1978}, which is valid in the Knudsen regime:
\begin{equation}
	\tau_{cond}^{-1} = \frac{3 \alpha f \rho_{sat} S_{max}}{2 a \rho_{ice}}\bigg(\frac{2 k T}{\pi m}\bigg)^{\frac{1}{2}}
\end{equation} 

where $a$ is the particule radius,  $\alpha$ is the molecular sticking coefficient, chosen to be equal to 0.7, which is similar to the 
mean H$_2$O sticking coefficient measured in the temperature range 190-235 K \citep{Skrotzki_2013}, 
as no measurement of the sticking coefficient of C$_6$H$_6$ ice was found in the literature. The $f$ parameter is equal to 3.8 and we used
a supersaturation maximum 10$^{-3} < S_{max} < $10$^{-1}$, as suggested by \cite{Rossow_1978}. $\rho_{sat}$ is the saturation C$_6$H$_6$ vapor density.

We focus here on the nadir observation at 87$^{\circ}$S of May 2013, where the C$_6$H$_6$ gas saturates at 0.025 mbar ($\sim$285 km) and 
138 K and derive $\tau_{cond} = a \times 10^{6\pm1}$ s with $a$ in $\mu$m.
If we consider that C$_6$H$_6$ condensation occurs over the saturated benzene gas scale height, i.e. $\sim$0.3 pressure scale height 
here, then by equalizing $\tau_{cond}$ and $\tau_{fall}\times$0.3, we obtain a range of 0.1 $\mu$m $< a <$ 1.7 $\mu$m with a mean equivalent radius
of 0.5 $\mu$m.  
This is consistent with the observational constraints on the equivalent radius that we determined from the fits of the C$_6$H$_6$ 
ice band (Section \ref{section_spectre_particl_shape}).
This estimated equivalent radius is also in agreement with the mean size of C$_6$H$_6$ ice particles of \cite{Barth_2017}'s 
microphysics model, which focus on 
equatorial conditions, where C$_6$H$_6$ condenses much deeper (around 90 km) than what we observe here.

Coagulation and coalescence processes at the C$_6$H$_6$ saturation pressure level (0.025 mbar), where the estimated ice particle number density 
is about 0.5 cm$^{-3}$, are negligible as their corresponding timescales, determined from Eqs. (33) and (37) of \cite{Rossow_1978} respectively, are two orders of magnitude higher than the sedimentation and condensation timescales.

By conservation of the mass flux we can estimate the mass mixing ratio of C$_6$H$_6$ ice ($q_{ice}$) from the gas mass mixing ratio at the 
condensation level ($q_{sat}$) with $q_{ice} \sim \frac{\tau_{sed}}{\tau_{sed}+\tau_{dyn}} q_{sat}$. 
From the estimated range of the C$_6$H$_6$ ice particle radius, $\tau_{sed}$ varies from 6.0 $\times$ 10$^{5}$ s to
7.3 $\times$ 10$^{5}$ s. This implies 0.05 $< q_{ice}/q_{sat} <$ 0.4, which is compatible within error bars with our derived
results in May 2013 and March 2015 if we extrapolate the ice mass mixing ratios of May 2013 toward the condensation level or if we 
extrapolate the gas mass mixing ratio of March 2015 to the saturation level (see Figures \ref{mixing_ratios_nadir} and \ref{mixing_ratios_limb}).

\subsection{Tentative identification of other ice signatures in the 680-710 cm$^{-1}$ range}

As mentioned in Section \ref{section_results}, unidentified signatures are observed in limb spectra at 79$^{\circ}$S near 687 cm$^{-1}$ 
and 702 cm$^{-1}$ in March 2015 (Figure \ref{spe_FP3_limb}) at 168, 205 and 239 km. 
Nadir spectra of May 2013 seem to display a common signature in the 695-700 cm$^{-1}$ spectral range that is more
prominent at 78$^{\circ}$S and 73$^{\circ}$S.
These signatures are $\sim$5-10 cm$^{-1}$ large and are detected at relatively 
low altitude as they are only observed on the 278-km limb spectrum and deeper and in nadir spectra, which probe the 5-20 mbar (150-100 km) region. 
This suggests possible ices as spectroscopic candidates.

\subsubsection{CH$_3$CN and other nitrile candidates}

We searched for possible ice candidates that could reproduce the observed limb spectra and that could condense at high altitude. 
Several nitrile ices display spectral signatures in the 650 - 800 cm$^{-1}$ spectral range. \cite{Moore_2010}
determined optical constants of several nitrile ices relevant for Titan's atmosphere. 
Their best candidates for our observed unidentified signature near 695 cm$^{-1}$ is CH$_3$CN. In order to verify if this nitrile ice could reproduce the 
observed limb spectra, we used their optical constants to calculate 
extinction cross sections for spherical particles of radius varying from 0.01 to 1.0 $\mu$m. 
The CH$_3$CN 695 cm$^{-1}$ ice band could contribute to the 695 cm$^{-1}$ emission feature but it should also display a stronger emission band at 773 cm$^{-1}$
that should be detectable in CIRS limb spectra if the observed 695 cm$^{-1}$ band intensity is reproduced. But, as we do not detect the 773 cm$^{-1}$ CH$_3$CN ice 
feature, we can exclude a contribution of this pure ice to the CIRS observed limb spectra. 
Therefore, CH$_3$CN pure ice does not explain the observed signal at 695 cm$^{-1}$. The other nitriles
investigated in \cite{Moore_2010}'s study (HC$_3$N, C$_2$H$_5$CN, HCN, 
and C$_2$N$_2$) do not have any spectral signature close to 700 cm$^{-1}$.

\subsubsection{The C$_2$H$_3$CN \textcolor{black}{potential} candidate}

\begin{figure}[p] 
	\centering
	
         \includegraphics[scale=0.7]{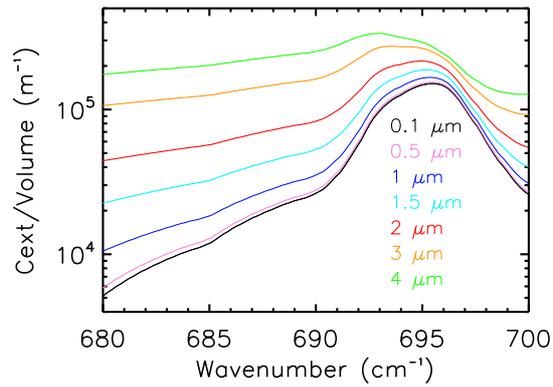}
	 \caption{\textcolor{black}{Spectral dependence of the extinction cross section per unit particle volume calculated for spherical
	 particules composed of pure C$_2$H$_3$CN ice with different radii from 0.1 $\mu$m (black line) to 4 $\mu$m (green).} }
	 \label{cross_section_C2H3CN}
	\end{figure}

\begin{figure}[p] 
	\centering
	 \includegraphics[scale=0.7]{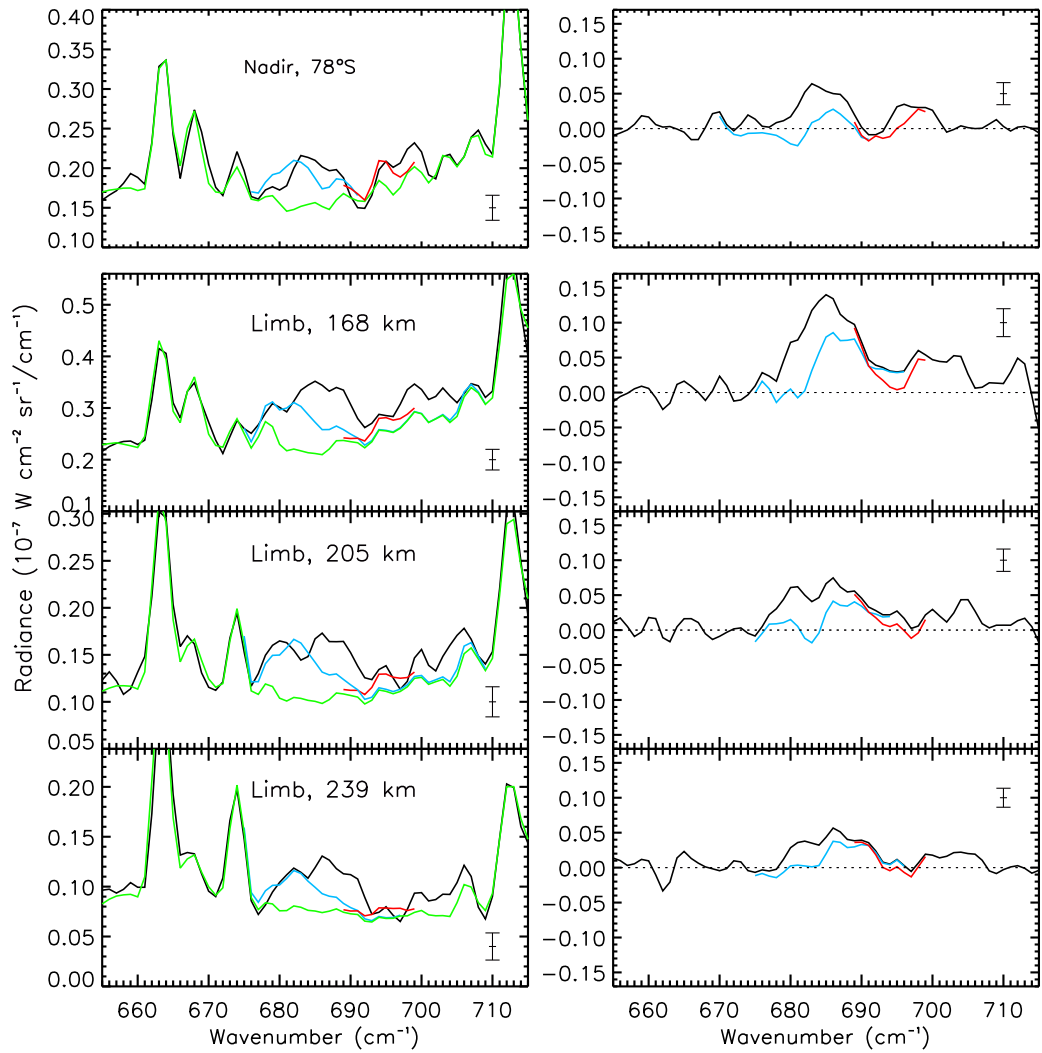}

	 \caption{Top panel: same as Figure \ref{spe_FP3} but only for the nadir spectrum at 78$^{\circ}$S and including the 
	 C$_2$H$_3$CN ice signature in red and C$_6$H$_6$ ice in blue. Lower panels: same as Figure \ref{spe_FP3_limb} but 
	 only for the limb spectra at 168, 205 and 239 km 
	 and including the C$_2$H$_3$CN ice signature in red and C$_6$H$_6$ ice in blue.}
	 \label{spe_C2H3CN}
	\end{figure}

\begin{figure}[p] 
	\centering

         \includegraphics[scale=0.5]{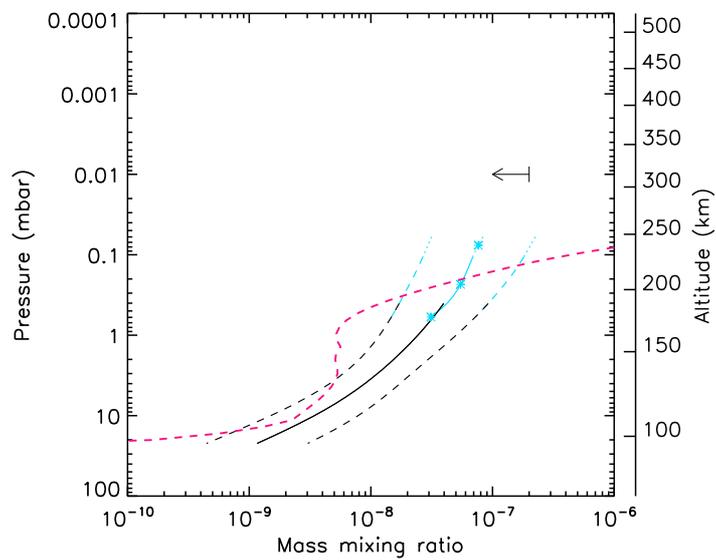}
    
	 \caption{Retrieved C$_2$H$_3$CN ice mass mixing ratio profile from the limb spectra (cyan line) and from the nadir spectrum
	 at 78$^{\circ}$S (black). Solid lines give the mean mass mixing ratio, while dashed lines give the 
	 1-$\sigma$ error bars. \textcolor{black}{Mean altitudes of the line-of-sight of the limb spectra are displayed as crosses.}
	 The pink dashed line gives the liquid-gas transition curve in Titan's atmosphere derived from \cite{Lide_2009}.
	 \textcolor{black}{Arrow represents upper limit of the C$_2$H$_3$CN gas mixing ratio derived from CIRS limb observation at 0.01 mbar.}}	
	\label{mixing_ratios_C2H3CN}
	\end{figure}

\cite{Russo_Khanna_1996} derived optical constants of acrylonitrile (C$_2$H$_3$CN) ice, which displays a single signature at 695 cm$^{-1}$
in the mid-IR range observed by CIRS. We utilized their optical constants to derive from Mie calculation the spectral dependence of the
 extinction cross sections of C$_2$H$_3$CN ice for spherical particles with 
radii varying from 0.01 to 4.0 $\mu$m \textcolor{black}{(see Figure \ref {cross_section_C2H3CN})}. 
\textcolor{black}{The noise level in nadir and limb spectra and the spectral contribution of the unknown spectral 
feature at 687 cm$^{-1}$ in limb spectra prevented us from deriving constraints on the size of the potential C$_2$H$_3$CN
ice particules. We then chose to perform} retrievals with the extinction cross section calculated for a particle radius of 0.5 $\mu$m (like for C$_6$H$_6$ ice).
Resulting fits of the 78$^{\circ}$S nadir spectrum of May 2013 and the March 2015 limb spectra are displayed in Figure \ref{mixing_ratios_C2H3CN} (a). 
The 695 cm$^{-1}$ signature is seen on limb spectra at 168 and 205 km, and possibly 239 km and we therefore applied a cutoff on the a priori optical depth profile of the C$_2$H$_3$CN ice at 0.05 mbar ($\sim$ 250 km). The retrieved 
C$_2$H$_3$CN ice mass mixing ratio profile from the 168, 205 km and 239 km limb spectra is displayed in Figure \ref{mixing_ratios_C2H3CN} (cyan line) and probe the 0.1 - 1 mbar region. This vertical profile was then used to model the nadir 695 cm$^{-1}$ 
signature at 78$^{\circ}$S. The resulting fit is displayed in Figure \ref{spe_C2H3CN} (top panel), 
and the retrieved mass mixing ratio at 78$^{\circ}$S is displayed in Figure \ref{mixing_ratios_C2H3CN} (in black). The nadir 
spectrum probes the 0.4-20 mbar (205 - 80 km) region. At 78$^{\circ}$S, we derive a C$_2$H$_3$CN ice mass mixing ratio half that of 
C$_6$H$_6$ ice at 0.1 mbar and 10 times smaller at 10 mbar.  
Figure \ref{mixing_ratios_C2H3CN} additionally displays the C$_2$H$_3$CN liquid-gas transition curve in Titan's atmosphere derived from 
data listed in 
\cite{Lide_2009}, as no sublimation data exist in the literature. Nevertheless, the slope of the sublimation curve near the triple point, 
of which the temperature is 189.6 K \citep{Lide_2009}, is always steeper than the slope of the liquid-gas curve in a phase transition pressure-temperature diagram. 
We therefore expect the sublimation curve to be located at lower pressure levels than the 
liquid-gas one in Figure \ref{mixing_ratios_C2H3CN}. Therefore, C$_2$H$_3$CN should condense at a pressure level lower than $\sim$0.2 mbar, or altitude higher than $\sim$200 km 
(corresponding to the level where the C$_2$H$_3$CN ice mass mixing ratio crosses the liquid-gas saturation curve). This is therefore 
compatible with the possible detection of C$_2$H$_3$CN ice emission feature at 239 km and below.

Acrylonitrile gas was detected by Cassini/INMS near 1000 km with volume mixing ratios varying from 3.5 $\times$10$^{-7}$ to 1$\times$10$^{-5}$ 
\citep{Vuitton_2007,Magee_2009, Cui_2009} \textcolor {black}{and from the Atacama Large Millimeter Array (ALMA) above 200 km
with mixing ratios comprised between 0.36 and 2.83$\times$10$^{-9}$ at 300 km \citep{Palmer_2017}}. 
If the 695 cm$^{-1}$ signature observed in limb spectra of March 2015 is due to acrylonitrile ice and if we assume that the retrieved ice
mass mixing ratio near 0.1 mbar is the one near the top of the  C$_2$H$_3$CN cloud (as the 695 cm$^{-1}$ signature is not detected at higher altitude), then because of mass flux conservation, the gas volume mixing ratio at $\sim0.1$ mbar should be higher than $\sim$4 $\times$10$^{-8}$, and also lower than the INMS values. 
There is currently no line list of the acrylonitrile gas in the 
mid-IR but \cite{Khlifi_1999} determined its IR spectrum, which displays many vibrational modes, and measured the corresponding integrated band intensities. In the mid-IR spectral region observed by CIRS, the three most intense acrylonitrile bands are located at 682, 953 and 972 
cm$^{-1}$.
The Q-branch of the $\nu_{14}$ bending mode of C$_2$H$_3$CN gas localized at 682 cm$^{-1}$ is superposed to the 682 cm$^{-1}$ spectral 
signature that we attributed here to the C$_6$H$_6$ ice $\nu_{4}$ bending mode. Nevertheless, we
can be sure that the $\nu_{14}$ C$_2$H$_3$CN gas Q-branch does not explain the 682 cm$^{-1}$ signature because this Q-branch is only 1-2 cm$^{-1}$ wide, 
which is not enough to reproduce the 5-10 cm$^{-1}$ wide band observed here. 
We can estimate the 682 cm$^{-1}$ C$_2$H$_3$CN Q-branch intensity using the \cite{Khlifi_1999} integrated intensity of 41 cm$^{-2}$atm$^{-1}$,
 the HC$_3$N or C$_6$H$_6$ gas spectral radiance observed by CIRS (at 663 and 674 cm$^{-1}$, respectively), and their integrated band intensities measured by 
\cite{Khlifi_1992b, Khlifi_1992}. To do so, we use the equation $\frac{I_1}{I_2} = \frac{q_1}{q_2} \frac{S_1}{S_2} \frac{B_1}{B_2}$, 
where $I_1$ and $I_2$ are observed radiances of species 1 and 2, respectively, $S_1$ and $S_2$ the band strengths for 
the corresponding absorption features,
$q_1$ and $q_2$ the volume mixing ratios and 
$B_1$ and $B_2$ the Planck radiance at the wavenumbers of the emission bands of species 1 and 2, respectively.
We applied a factor $\frac{1}{3}$ to the integrated band intensities of HC$_3$N and C$_6$H$_6$ in order to estimate their Q-branches 
integrated intensities.
If we assume a C$_2$H$_3$CN gas volume mixing ratio of 5$\times$10$^{-8}$ at 0.1 mbar ($\sim$250 km), then it would be reasonable to estimate a mixing ratio value of the order of 1$\times$10$^{-7}$ at 0.01 mbar ($\sim$320 km), which is the level above which we derive information for the molecular gas mixing ratios at limb viewing.
Using the Q-branches intensities of HC$_3$N and C$_6$H$_6$ from the 320-km limb spectrum of 
Figure \ref{spe_FP3_limb}, we then estimate that the radiance of the C$_2$H$_3$CN Q-branch would be 0.03$\times$10$^{-7}$ 
W cm$^2$ sr$^{-1}$/cm$^{-1}$, which is similar to the noise equivalent spectral radiance in CIRS spectrum (see residuals in Figure \ref{spe_FP3_limb}).
Doing the same calculation for the $\nu_{12}$ and $\nu_{13}$ C$_2$H$_3$CN Q-branches at 972 and 953 cm$^{-1}$, respectively, 
which have a total integrated intensity of 201 cm$^{-2}$atm$^{-1}$ (two Q-branches of $\sim$100 cm$^{-2}$atm$^{-1}$ each) give a predicted 
radiance 3 times smaller than the $\nu_{14}$ one, which is not detectable here. 
From the noise level ($\sim$0.02$\times$10$^{-7}$ W.cm$^2$.sr$^{-1}$/cm$^{-1}$) of the 320-km limb spectrum, the corresponding 3-$\sigma$
upper limit of the C$_2$H$_3$CN gas volume mixing ratio is 2$\times$10$^{-7}$ at 0.01 mbar at 79$^{\circ}$S in March 2015.
If C$_2$H$_3$CN gas displays a gradient similar to other molecules (C$_6$H$_6$, HC$_3$N, HCN, C$_2$H$_2$, C$_4$H$_2$ and 
CH$_3$C$_2$H), then we would expect an increase of its volume mixing ratio by a factor $\sim$10 from 0.01 to 1$\times$10$^{-4}$ mbar
where molecule volume mixing ratios reach values comparable to INMS measurements at 1000 km. Thus, this would correspond 
to a C$_2$H$_3$CN gas volume mixing ratio consistent with the INMS measurements. Such a vertical volume mixing ratio profile would then 
be consistent with a non detection of the C$_2$H$_3$CN gas emission in CIRS spectra, while its ice emission could be detected.

\subsubsection{The 687 cm$^{-1}$ signature}

We did not find in the literature any pure ice candidate that could explain the 687 cm$^{-1}$ spectral feature. 
It is possible that a complex vibrational interaction with aerosols and/or other ices produces a shift in the frequency of
the ice spectral signature and/or modifies the shape of the spectral signature (see for instance the case of predicted
H$_2$O-CO$_2$ core-shell particles extinction spectra of \cite{Isenor_2013}). Ongoing laboratory experiment efforts 
on identifying the ice mixtures that could reproduce this spectral feature are conducted at NASA/GSFC 
\citep{Anderson_2017}. 

\subsubsection{Voyager observations}

It is interesting to notice that Voyager limb observations of the north pole of Titan, during the northern winter, displayed 
several unidentified spectral signatures \citep{Coustenis_1999}, and in particular the 682 cm$^{-1}$ signature that we 
detect here in CIRS spectra at the south pole in the middle of fall. \cite{Coustenis_1999} also detected a large spectral signature 
in the 690-710 cm$^{-1}$ that could correspond to the 702 cm$^{-1}$ unidentified signatures observed here.
Additionally, the 700 cm$^{-1}$ spectral signature was observed in CIRS spectra in the northern polar region during winter and we therefore 
may be seeing here the appearance of this signature that will remain during the entire southern winter.

\section{Conclusion and perspectives}

This study reports the first detection of C$_6$H$_6$ ice in Titan's atmosphere from its $\nu_{4}$ C-H bending mode at 682 cm$^{-1}$. 
We have constrained the vertical profile of the benzene ice cloud in March 2015 at 79$^{\circ}$S and derived its spatial distribution 
from the south pole to 68$^{\circ}$S in May 2013. Top of the C$_6$H$_6$ ice cloud is observed deeper with increasing distance 
from the south pole, while the ice mass mixing ratio seems to be constant at $\sim$10 mbar from one latitude to another. 
We derived an upper limit of 1.5 $\mu$m for the equivalent 
radius of C$_6$H$_6$ ice particles from the shape of the 682 cm$^{-1}$ emission band, which is in agreement with our estimation of the 
ice particle size from condensation growth and falling timescales comparison.
%Nadir spectra observed at latitudes from 73$^{\circ}$S to 68$^{\circ}$S, at the edge of the southern polar cloud where the top of the cloud is 
%observed deeper, are better fitted with inhomogeneous particles made of a haze core and a pure C$_6$H$_6$ ice shell.
Observation of the C$_6$H$_6$ ice spectral signature is associated with unidentified spectral signatures observed in limb spectra 
in the 685-710 cm$^{-1}$ spectral range in the 0.1-1 mbar region and in the 78$^{\circ}$S and 73$^{\circ}$S nadir spectra observed in 
May 2013. We tested several pure nitrile ices that have emission bands in this range and with available optical constants. We derived 
that C$_2$H$_3$CN would be the only pure nitrile ice candidate \textcolor{black}{(we also tested HCN, HC$_3$N,CH$_3$CN, C$_2$H$_5$CN and C$_2$N$_2$)}
that could contribute in this spectral region with a band centered at 695 cm$^{-1}$, from which we 
derived an ice mass mixing ratio between 2 and 10 times smaller than the C$_6$H$_6$ ice one in the 0.1 - 10 mbar region. 
We inferred an upper limit of $\sim 2 \times$10$^{-7}$ for the C$_2$H$_3$CN gas volume mixing ratio at 0.01 mbar at 79$^{\circ}$S in March 2015.

Even if the present paper focuses on the study of two CIRS datasets of the south pole in May 2013, from nadir viewing, and in March 2015 from 
limb geometry, we observed the 682 cm$^{-1}$ spectral signature in many other southern polar nadir spectra 
from early 2013 to early 2015. The C$_6$H$_6$ ice emission band presents some temporal variations and the 695 cm$^{-1}$ spectral 
feature,\textcolor{black} {that we tentatively attribute} to the C$_2$H$_3$CN ice, is also observed in other data. 
A detailed study of the south polar CIRS nadir data and limb spectra acquired after March 2015 will allow us to derive more information on 
the formation, evolution and composition of the 
southern stratospheric polar cloud that appeared during the southern fall of Titan.

\section{Acknowledgment}

This work was funded by the French Centre National d'Etudes 
Spatiales and the Programme National de Plan\'etologie (INSU).
We thank E. Barth and C. Anderson for helpful discussions.

\bibliographystyle{elsarticle-harv}
\bibliography{ma_biblio.bib}

%\nocite{Vinatier_2010b}
%\nocite{Vinatier_2010a}

%% Authors are advised to submit their bibtex database files. They are
%% requested to list a bibtex style file in the manuscript if they do
%% not want to use elsarticle-harv.bst.

%% References without bibTeX database:

%% \begin{thebibliography}{00}

%% \bibitem must have one of the following forms:
%%   \bibitem[Vinatier et al.(2010b)]{Vinatier_2010b}
%%   \bibitem[Jones et al.(1990)Jones, Baker, and Williams]{key}...
%%   \bibitem[Jones et al., 1990]{key}...
%%   \bibitem[\protect\citeauthoryear{Jones, Baker, and Williams}{Jones
%%       et al.}{1990}]{key}...
%%   \bibitem[\protect\citeauthoryear{Jones et al.}{1990}]{key}...
%%   \bibitem[\protect\astroncite{Jones et al.}{1990}]{key}...
%%   \bibitem[\protect\citename{Jones et al., }1990]{key}...
%%   \harvarditem[Jones et al.]{Jones, Baker, and Williams}{1990}{key}...%%
%
%% \bibitem[ ()]{}
%
%%% \end{thebibliography}
%
%

\newpage

\begin{center}
\begin{singlespace}
\begin{table}
     \begin{footnotesize}
   \begin{tabular}{c c c c c c c c}
      \hline
         \small{ } &\small{\textbf{$\theta$}} & \small{\textbf{longitude}}  &  \small{\textbf{number of }}    &   \small{\textbf{mean}}  & \small{\textbf{Cassini distance}} &     \small{\textbf{Spatial}}\\

         \small{ } &\small{\textbf{range}}    &\small{\textbf{range}}       &  \small{\textbf{spectra}}    &  \small{\textbf{$\theta$ ($^\circ$)} }  &  \small{\textbf{to surface} } & \small{\textbf{resolution} } \\

         \small{ } &\small{\textbf{($^\circ$)}}    &\small{\textbf{($^{\circ}$ W})}       &  \small{\textbf{}}    &  \small{  }   &  \small{ \textbf{(km)} } & \textbf{($^{\circ}$ latitude)}  \\

      \hline

  \small{24th May 2013}  \\

  \small{\textbf{90$^{\circ}$S - 85$^{\circ}$S}}  \\
    
        \small{FP3} &\small{39.2 - 48.0}    &\small{6.9 - 219.2}       &  \small{17}  &   \small{43.6}   &  \small{361742} &  \small{2.19}  \\
      
        \small{FP4} &\small{39.8 - 48.7}    &\small{56.1 - 353.6}      &  \small{19}   &   \small{44.2}   &  \small{361483} & \small{2.20} \\

  \small{\textbf{85$^{\circ}$S - 80$^{\circ}$S}}  \\
    
        \small{FP3} &\small{34.8 - 49.1}    &\small{139.2 - 352.0}     &  \small{29}    &   \small{40.9}   &  \small{360239} &  \small{2.19}\\
      
        \small{FP4} &\small{34.6 - 50.6}    &\small{131.0 - 348.9}     &  \small{27}    &   \small{40.2}   &  \small{360700} & \small{2.19} \\

  \small{\textbf{80$^{\circ}$S - 75$^{\circ}$S}}  \\
    
        \small{FP3} &\small{29.7 - 53.7}    &\small{134.0 - 359.4}     &  \small{49}    &  \small{38.4}   &  \small{360625} &  \small{2.19}\\
      
        \small{FP4} &\small{29.6 - 52.8}    &\small{137.2 - 350.9}     &  \small{41}    &   \small{38.1}   &  \small{360854} &  \small{2.19}\\

  \small{\textbf{75$^{\circ}$S - 70$^{\circ}$S}}  \\
    
        \small{FP3} &\small{24.4 - 56.9}    &\small{133.4 - 358.0}     &  \small{76}    &  \small{36.5}   &  \small{360811} & \small{2.19} \\
      
        \small{FP4} &\small{24.5 - 55.9}    &\small{134.6 - 356.8}     &  \small{65}    &   \small{37.1}   &  \small{360333} & \small{2.19} \\

  \small{\textbf{70$^{\circ}$S - 65$^{\circ}$S}}  \\
    
        \small{FP3} &\small{20.2 - 39.3}    &\small{186.5 - 324.3}     &  \small{58}    &  \small{28.5}   &  \small{360568}  & \small{2.19} \\
      
        \small{FP4} &\small{20.5 - 39.7}    &\small{187.2 - 324.0}     &  \small{51}    &   \small{29.2}   &  \small{360219} & \small{2.18} \\

  \small{\textbf{65$^{\circ}$S - 60$^{\circ}$S}}  \\
    
        \small{FP3} &\small{20.3 - 39.8}    &\small{187.9 - 325.0}     &  \small{59}    &  \small{29.6}   &  \small{358945} &  \small{2.18} \\
      \small{2.18} 
        \small{FP4} &\small{20.0 - 39.6}    &\small{187.6 - 324.7}     &  \small{56}    &   \small{29.8}   &  \small{355775} & \small{2.16}  \\

   \small{\textbf{55$^{\circ}$S - 50$^{\circ}$S}}  \\
    
        \small{FP3} &\small{4.2 - 19.8}    &\small{225.2 - 283.9}     &  \small{163}    &  \small{10.1}   &  \small{405761} &  \small{2.46} \\
      
        \small{FP4} &\small{4.3 - 19.6}    &\small{225.4 - 284.1}     &  \small{45}    &   \small{11.9}   &  \small{362164} &  \small{2.20}\\

   \small{10th December 2014}  \\ 
   
       \small{\textbf{80$^{\circ}$S - 75$^{\circ}$S}}  \\
    
        \small{FP3} &\small{44.7 - 60.6}    &\small{5.6 - 357.8}     &  \small{40}    &  \small{51.3}   &  \small{326030-306569} &  \small{1.86 -1.98}\\
      
        \small{FP4} &\small{44.5 - 61.6}    &\small{8.2 - 357.6}     &  \small{35}    &   \small{52.0}   &  \small{325484-307112} &  \small{1.86 - 1.98}\\
     
      \hline
     
  \end{tabular}
 \end{footnotesize}    
      \caption{Characteristics of the averaged nadir spectra acquired at a spectral resolution of 2.8 cm$^{-1}$. $\theta$ is the 
      emission angle. Spectra were extracted from the 4.3.1 version of the CIRS database. Latitudes are those corresponding to the 
      solid body latitudes extracted from the CIRS database.}

    \end{table}   
   \end{singlespace} 
   \end{center}

\newpage 

\begin{center}
\begin{singlespace}
\begin{table}
     \begin{footnotesize}
   \begin{tabular}{c c c c c c}
      \hline
   
      \small{\textbf{Focal}} &\small{\textbf{Mean}} & \small{\textbf{Mean}}  &  \small{\textbf{Mean vert.}}  &   \small{\textbf{Altitude}}  & \small{\textbf{Number of averaged}} \\

      \small{\textbf{plane}} &\small{\textbf{lat.}} & \small{\textbf{long.}}  &  \small{\textbf{resolution (km)}}  &   \small{\textbf{range (km)}}  & \small{\textbf{spectra}}  \\

      \hline

        \small{FP4} &\small{79$^{\circ}$}   &\small{135$^{\circ}$ W}   &  \small{42}  &   \small{123 - 549}   &  \small{53 - 134}  \\
      
        \small{FP3} &\small{81$^{\circ}$}   &\small{137$^{\circ}$ W}   &  \small{42}  &   \small{128 - 548}   &  \small{50 - 124} \\

      \hline
     
  \end{tabular}
 \end{footnotesize}    
      \caption{Characteristics of the averaged limb spectra acquired at a spectral resolution of 0.5 cm$^{-1}$ on 16 March 2015 (flyby
      T110, solar longitude of 66$^{\circ}$). The first four columns give the focal plane, the mean latitude, longitude and vertical resolution 
               of the averaged limb spectra. Fifth column gives the minimum and maximum altitudes of the line-of-sight of averaged spectra. 
	       Last column gives the minimum and maximum number per averaged spectra. 
            Spectra were extracted from the 4.3.1 version of the CIRS database. }

    \end{table}   
   \end{singlespace} 
   \end{center}

\newpage

\newpage

\newpage

\newpage

\newpage 

\newpage 

\newpage

\newpage

\newpage

\end{document}